\documentclass[journal]{IEEEtran}

\newtheorem{theorem}{Theorem}
\newtheorem{proposition}{Proposition}

\usepackage{booktabs}
\usepackage{bm}
\usepackage{graphicx}
\usepackage[cmex10]{amsmath}
\usepackage{cite}
\usepackage{amssymb}
\usepackage{subfigure}
\usepackage{extarrows}
\usepackage[letterspace = -2]{microtype}
\usepackage{algorithm}
\usepackage{algpseudocode}
\usepackage{algpascal}
\usepackage{float}

\usepackage{color}
\usepackage{xcolor}

\newfloat{algorithm}{t}{lop}

\begin{document}

\title{Waveform Design for Wireless Power Transfer with Limited Feedback}
\author{
Yang~Huang
and Bruno~Clerckx,~\IEEEmembership{Senior Member,~IEEE}
\thanks{Y. Huang and B. Clerckx are with the Department of Electrical and Electronic Engineering, Imperial College London, London SW7 2AZ, United Kingdom (e-mail: \{y.huang13, b.clerckx\}@imperial.ac.uk). Yang's work was supported by the China Scholarship Council Imperial Scholarship. This work has also been partially supported by the EPSRC of UK, under grant EP/P003885/1.}
}

\maketitle

\vspace{-0.2cm}
\begin{abstract}
Waveform design is a key technique to jointly exploit a beamforming gain, the channel frequency-selectivity and the rectifier nonlinearity, so as to enhance the end-to-end power transfer efficiency of Wireless Power Transfer (WPT). Those waveforms have been designed assuming perfect channel state information at the transmitter. This paper proposes two waveform strategies relying on limited feedback for multi-antenna multi-sine WPT over frequency-selective channels. In the waveform selection strategy, the Energy Transmitter (ET) transmits over multiple timeslots with every time a different waveform precoder within a codebook, and the Energy Receiver (ER) reports the index of the precoder in the codebook that leads to the largest harvested energy. In the waveform refinement strategy, the ET sequentially transmits two waveforms in each stage, and the ER reports one feedback bit indicating an increase/decrease in the harvested energy during this stage. Based on multiple one-bit feedback, the ET successively refines waveform precoders in a tree-structured codebook over multiple stages. By employing the framework of the generalized Lloyd's algorithm, novel algorithms are proposed for both strategies to optimize the codebooks in both space and frequency domains. The proposed limited feedback-based waveform strategies are shown to outperform a set of baselines, achieving higher harvested energy.
\end{abstract}


\section{Introduction}
With the evolution of the Internet of Things, sensors and low-power devices become numerous and smaller. They also might be deployed in unreachable or hazard environment, such that battery replacement becomes inconvenient. An issue arises as how to power these sensors and low-power devices in a reliable, controllable, cost-effective and user-friendly manner. Radiative Wireless Power Transfer, referred to as WPT, has been regarded as a promising solution, where an Energy Receiver (ER) exploits rectennas to convert Radio-Frequency (RF) signals sent by an Energy Transmitter (ET) into DC power.
Aside optimizing the rectenna circuit, another promising approach to enhance the end-to-end power transfer efficiency is to design efficient WPT signals (including waveforms, beamforming and power allocation) \cite{ZCZ16}. It was observed through RF measurements in the RF literature that the RF-to-DC conversion efficiency is a function of the input waveforms and can be enhanced by a superposition of sinewaves over frequencies with uniform frequency spacing\cite{BBFCGC15, CM16, BCCG13, BC11}.
Following this observation, the first systematic analysis, design and optimization of waveforms for WPT was conducted in \cite{CB16}. Those waveforms are adaptive to the Channel State Information (CSI) and jointly exploit a beamforming gain, the frequency-selectivity of the channel and the rectifier nonlinearity so as to maximize the amount of harvested DC power. Since then, further enhancements have been made to waveform optimization adaptive to CSI with the objective to reduce the complexity of the design and extend to multi-user setup and large scale multi-antenna multi-sine WPT architecture \cite{HC16SPAWC, HC16TSParxiv, CB17arXiv, MZZ17arxiv}.

One essential feature of the existing waveform literature \cite{ZCZ16, CB16, HC16SPAWC, HC16TSParxiv, CB17arXiv} is to account for the nonlinearity of the rectifier, where the nonlinear rectification process is characterized by truncated power series models. These nonlinear rectenna models are derived from the Shockley diode equation \cite{BCCG13} and truncated to at least the 4th-order term\cite{BC11}. Though the conventional linear rectenna model\cite{ZZH13} is also a power series model derived from the diode equation, the model is truncated to the 2nd-order term. However, for multi-sine WPT, if the average input power into the rectenna is between -30dBm and 0 dBm, the contribution of the 4th-order term to the rectenna output cannot be neglected\cite{ZCZ16}, and the nonlinear rectenna model should be exploited for waveform designs. Circuit simulations in \cite{CB16, CB17arXiv} validate the 4th-order truncation (nonlinear) model and confirm the inaccuracy of the linear model.

Acquiring the CSI at the Transmitter (CSIT) is shown in \cite{CB16} to be very useful to boost the rectenna DC output power. Despite the recent progress and promising gains, a major limitation of the current waveform design literature is that perfect CSIT has been assumed.
In this paper, we further explore waveform design, but consider codebook-based waveform strategies that account for a communication link with limited feedback between the ER and the ET.
To optimize these waveform codebooks, we develop novel codebook design algorithms, by employing the framework of the Generalized Lloyd's Algorithm (GLA)\cite{LBG80}.
GLA is originally used to generate codebooks for Vector Quantization (VQ)\cite{LBG80}, where for a given training set, GLA alternatively optimizes partition of the set and the centroid (which is a quantizer codeword) of each partition cell.
%
Additionally, GLA has been widely used to design beamforming codebooks for Multiple-Input Multiple-Output (MIMO) communications\cite{XG06, KJL12, JWSWSC12, CKK08}. To alleviate codebook search complexity, GLA is also used to transform a Random Vector Quantization (RVQ) codebook \cite{CO13} into a Tree-Structured (TS) codebook\cite{SM11}.
In the GLA for VQ or MIMO communications, optimizing a codeword for a given partition cell boils down to a Rayleigh quotient maximization, whose global optimum is a dominant eigenvector\cite{XG06}.
Due to the nonlinear rectenna model, our waveform codeword optimization turns out to be a quartic optimization, which is NP-hard, and cannot be solved by the GLAs for VQ or MIMO communications.

Spatial domain energy beamforming for narrowband single-frequency WPT relying on channel acquisition has been investigated\cite{YHG14, ZZ15, XZ14, CKC17, XZ16}. In these schemes, channel acquisition approaches can be classified into three categories:
1) An ER estimates the channel and returns quantized CSI to an ET\cite{YHG14}. However, the ER may suffer from stringent energy and hardware constraints and be unable to perform channel estimation.
2) The ET estimates the channel by processing pilot signals sent by the ER\cite{ZZ15, ZZjun15}, where it is assumed that the ER has enough power to transmit pilot signals before Energy Harvesting (EH). Hence, this method may be inapplicable to an ER suffering from a stringent energy constraint.
3) The ET transmits training signals, receives limited feedback from the ER and estimates the channel, where the feedback reflects information on the amount of power harvested by the ER\cite{XZ14, CKC17, XZ16}.
In \cite{XZ14}, by performing an algorithm derived from the Analytical Center Cutting Plane Method (ACCPM), the ET updates the channel estimate and adjusts the training signals relying on a feedback bit from the ER. The feedback indicates an increase or decrease in power harvested from adjacent training signals.
In \cite{CKC17}, the received power-based estimation updates the channel estimate by Least Square Estimation (LSE) or a Kalman filter algorithm. The amount of power harvested at the ER is quantized and fed back to the ET for the update.
In \cite{XZ16}, the ACCPM-based channel acquisition \cite{XZ14} is generalized to a received power-based estimation.
The limitations of those works are that the linear model for the rectenna has been assumed and that the signal processing employed by \cite{XZ14, CKC17, XZ16} are inapplicable to the nonlinear model-based WPT. The ACCPM employed by \cite{XZ14,XZ16} is only applicable to convex optimizations, while the nonlinear model leads to nonconvex problems. The LSE and the Kalman filter algorithm in \cite{CKC17} are only applicable to the EH model, where harvested power is a linear function of an energy channel matrix plus noise.

In contrast to state-of-the-art WPT waveform designs that assume perfect CSIT, this paper relaxes the assumption on CSIT and considers that an ET determines the preferred waveform for multi-antenna multi-sine WPT according to limited feedback sent by an ER. As the ER may suffer from stringent power and hardware constraints, the feedback only carries information on the rectenna output voltage $v_\text{out}$. Although spatial beamforming relying on limited feedback has been designed for narrowband WPT\cite{XZ14, CKC17, XZ16}, this paper optimizes waveforms for multi-antenna multi-sine WPT (therefore encompassing both space and frequency domains). In order to reduce the signal processing time for an ET and come up with limited feedback-based waveform strategies for multi-antenna multi-sine WPT, we propose waveform strategies that generate waveforms from offline designed codebooks. Our contributions are listed as follows.

\begin{figure}[!t]
\centering
\includegraphics[width = 3.4in]{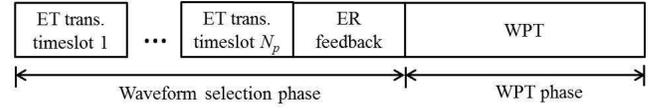}
\caption{Time frame for the waveform selection-based WPT.}
\label{FigBeamSelectionProtocol}
\end{figure}
\begin{figure}[!t]
\centering
\includegraphics[width = 3.4in]{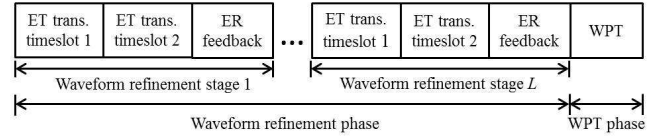}
\caption{Time frame for the waveform refinement-based WPT.}
\label{FigBeamTrainProtocol_ColocUsers}
\end{figure}
\textit{First}, we propose WPT strategies based on Waveform Selection (WS) and Waveform Refinement (WR), respectively.
During the WS phase as shown in Fig. \ref{FigBeamSelectionProtocol}, the ET essentially conducts an exhaustive search over an $N_p$-codeword codebook of waveform precoders, by sending a multi-sine signal generated by a codeword (i.e. a precoder) in each ET transmission timeslot. After the ET transmission timeslots, the ER feeds back the index of the waveform precoder that maximizes $v_\text{out}$.
Unlike the WS procedure, the WR procedure as shown in Fig. \ref{FigBeamTrainProtocol_ColocUsers} successively refines waveforms at each WR stage, by conducting searches over subcodebooks in a Tree-Structured (TS) waveform codebook. That is, multi-sine signals transmitted in timeslots 1 and 2 at WR stage $l$ are respectively generated by two codewords in a subcodebook that has higher resolution than the subcodebook used at stage $(l-1)$. The ER one-bit feedback at stage $(l-1)$ determines which subcodebook would be used at stage $l$, where the feedback bit indicates an increase or decrease in $v_\text{out}$ during timeslots 1 and 2 at WR stage $(l-1)$.
If the highest level of the TS codebook has $N_p$ codewords, $2 \log_2 N_p$ energy signals are sent during the WR phase.

\textit{Second}, in order to lay the basis of codebook design for the WS/WR-based WPT, we address the problem of solving a Sample Average Approximation (SAA) problem, which approximates the problem of maximizing the expected value of $v_\text{out}$ by averaging a sample of realizations. The SAA problem turns out to be a quartic optimization, which is nonconvex and NP-hard. To solve the SAA problem, we propose an iterative SAA algorithm, which converges to a Karush-Kuhn-Tucker (KKT) point (but not necessarily a global optimum) of the SAA problem. It is shown that the computational complexity at each iteration is not a function of the number of realizations in the sample for approximation.

\textit{Third}, we develop an algorithm to derive the waveform codebook for the WS-based WPT. Inspired by the GLA, we decouple optimizations over the training set partition and the codewords. However, as the SAA algorithm only converges to a KKT point, solving the codewords optimization subproblem by the SAA algorithm does not necessarily yield a global optimum\footnote{This is different from the GLAs for VQ and MIMO communications, where the codewords optimization always yields a global optimum, which guarantees the convergence of the alternating optimization of the codewords and the training set partition.}. In order to establish the convergence of the codebook design algorithm, we design a specific initial point for the SAA algorithm. The optimized codebook converges to a KKT point of the codewords subproblem.

\textit{Fourth}, we derive a TS codebook design algorithm for the WR-based WPT, based on the codebook design algorithm for the WS-based WPT. We reveal that although the WR-based WPT can reduce the overhead caused by searching out a preferred waveform, waveforms generated from a TS codebook may yield a lower average $v_\text{out}$ than those generated from the codebook for the WS-based WPT.

\textit{Fifth}, important observations are made from the simulations.
1) The proposed waveform strategies relying on limited feedback can outperform a set of baselines, achieving a higher average $v_\text{out}$.
2) The average $v_\text{out}$ of the WS-based WPT scales with an increasing number of sinewaves. Increasing the number of feedback bits enables a faster increase in $v_\text{out}$ as a function of the number of sinewaves.
3) Although a large bandwidth leads to a higher average $v_\text{out}$ in the presence of perfect CSIT, multi-sine WPT relying on limited feedback can benefit from a small bandwidth, achieving a higher average $v_\text{out}$, in the presence of a small number of feedback bits.
4) The proposed waveform strategies, though based on limited feedback, can significantly outperform the linear model-based waveform design based on perfect CSIT, in terms of the average $v_\text{out}$.
5) The tradeoff between the duration of the WS (or WR) phase and the duration of the WPT phase is investigated. It is shown that though the WS strategy outperforms the WR strategy in terms of harvested energy measured during the WPT phase (as shown in Figs. \ref{FigBeamSelectionProtocol} and \ref{FigBeamTrainProtocol_ColocUsers}), WR is preferred over WS when the overhead of searching a preferred waveform is accounted for.

\emph{Organization:} Section \ref{SecSystemModel} introduces the system model. Section \ref{SecMaxEwsumVout} proposes the SAA algorithm. Sections \ref{SecBeamSelectProtocol} and \ref{SecBeamTrainProtocol} respectively propose the WS and the WR-based WPT, whose performance is evaluated in Section \ref{SecSimResults}. Conclusions are drawn in Section \ref{SecConclu}.

\emph{Notations:} Matrices and vectors are in bold capital and bold lower cases, respectively. The notations $(\cdot)^T$, $(\cdot)^\star$, $(\cdot)^\ast$, $(\cdot)^H$, $\text{Tr}\{\cdot\}$, $\|\cdot\|$ and $|\cdot|$ represent the transpose, optimized solution, conjugate, conjugate transpose, trace, 2-norm and absolute value, respectively. The notation $\text{Card}\left(\mathcal{A}\right)$ represents the cardinality of the set $\mathcal{A}$; $\left[\mathbf{A}\right]_n$ denotes the $n$\,th column of $\mathbf{A}$; $\lceil x \rceil$ rounds the real $x$ to the nearest integer no less than $x$. In $\mathbf{h}_q^{[t]}$, the superscript $[t]$ is the index of the realizations of the random variable $\mathbf{h}_q$; while in $\mathbf{t}_q^{(l)}$, the superscript $(l)$ is the index of iterations, such that $\mathbf{t}_q^{(l)}$ means $\mathbf{t}_q$ in the $l$\,th iteration.

\section{System Model}\label{SecSystemModel}
\subsection{Signal Transmission and Rectification Models}\label{SecSignalTrans}
In the multi-antenna multi-sine WPT system, an ET equipped with $M$ antennas delivers multi-sine energy signals over $N$ frequencies to an ER equipped with $Q$ rectennas\footnote{Multi-sine signals are studied, as measurements have shown that they can significantly enhance the RF-to-DC power conversion efficiency\cite{BBFCGC15, CM16, BCCG13, BC11, CB16}. Furthermore, multi-sine signals are analytically tractable and widely used in major wireless systems e.g. OFDM in communications \cite{BBFCGC15, CO13}. Another interesting architecture for the ER is such that signals at antennas can be combined, e.g. \cite{OCV11}, and input into a common rectifier. Waveform strategies accounting for this ER architecture are left for future studies.}.
The channel gain between antenna $m$ and rectenna $q$ at frequency $n$ is $h_{q,(n-1)M + m} \in \mathbb{C}$.
Block fading channel model is considered, such that channel frequency responses remain constant over an entire time frame (as shown in Fig. \ref{FigBeamSelectionProtocol} or Fig. \ref{FigBeamTrainProtocol_ColocUsers}), which consists of a WS (or WR) phase and a WPT phase.
The complex version of the multi-sine signal transmitted by antenna $m$ at the ET is designated as $\tilde{x}_m(t) \triangleq \sum_{n = 1}^N s_{(n-1)M+m} e^{j\omega_n t}$, where the complex variable $s_{(n-1)M+m}$ collects the magnitude and the phase of the cosine signal at angular frequency $\omega_n$. Therefore, the transmitted RF signal by antenna $m$ can be expressed as $x_m(t) = \sqrt{2} \text{Re}\{\tilde{x}_m(t)\}$. The sinewaves are uniformly spaced, such that $\omega_n \!=\! \omega_1 \! + \! (n\!-\!1)\Delta_\omega$, for $n \! = \! 1, \ldots, N$ and $\omega_1 \!> \! (N \! - \! 1)\Delta_\omega/2$. The complex RF signal traveling through the channel between transmit antenna $m$ and rectenna $q$ can be written as
\begin{equation}
\label{EqYq_tilde}
\tilde{y}_{q, m}(t) = \textstyle{\sum_{n = 1}^N} s_{(n - 1)M+m} h_{q,(n-1)M+m} e^{j\omega_n t}\,.
\end{equation}
Hence, the RF signal transmitted by the $M$ antennas and input into rectenna $q$ at the ER can be formulated as
\begin{equation}
\label{EqYq}
y_q(t) = \sqrt{2} \text{Re}\left\{ \tilde{y}_q(t) \right\} = \sqrt{2} \text{Re}\left\{\textstyle{\sum_{m = 1}^M} \tilde{y}_{q,m}(t)\right\}\,.
\end{equation}
The multi-sine waveform $y_q(t)$ is then rectified by rectenna $q$, which outputs DC voltage $v_{\text{out},q}$. In the presence of multi-sine WPT, $v_{\text{out},q}$ is shown to be a nonlinear function of input waveforms $y_q(t)$\cite{BCCG13, BC11, BBFCGC15, CB16}. Hence, the nonlinear model constructed in \cite{HC16TSParxiv, HC16SPAWC} is employed in this paper. In order to obtain a tractable nonlinear model, the derivation of this nonlinear model follows the approach in \cite{Wetenkamp83} and assumes a low-power input signal and a very high impedance load, such that the rectenna output DC current is approximately equal to zero\cite{Wetenkamp83}. Then, by manipulating the Shockley diode equation, output DC voltage $v_{\text{out},q}$ of the $q$\,th rectenna can be expressed as a function of $y_q(t)$ \cite{HC16TSParxiv, HC16SPAWC}.
In the model, ideal matching network and low pass filter $f_\text{LPF}(\cdot)$ are assumed, such that the non-DC harmonics can be filtered. The model is given by
\begin{equation}
\label{EqVoutq_Scalar2}
v_{\text{out},q} = \beta_2 f_\text{LPF}\left(y_q^2(t)\right) + \beta_4 f_\text{LPF}\left(y_q^4(t)\right)\,,
\end{equation}
where $\beta_2 \triangleq R_\text{ant}/(2 n_\text{if} V_\text{T})$ and $\beta_4 \triangleq R_\text{ant}^2/(24 n_\text{if}^3 V_\text{T}^3)$; parameters $V_\text{T}$, $n_\text{if}$ and $R_{\text{ant}}$ stand for the thermal voltage of the diode, ideality factor (set as 1 for simplicity) and the antenna impedance (set as $50\Omega$) of the rectenna, respectively.
Note that the parameters $R_\text{ant}$, $n_\text{if}$ and $V_\text{T}$ are constants, but not functions of $y_q(t)$.
Due to the assumption of the ideal low pass filter, $v_{\text{out},q}$ remains constant and depends on the peak of the input waveform $y_q(t)$. That is, $v_{\text{out},q}$ increases with the increasing peak of $y_q(t)$. However, the nonlinear model (\ref{EqVoutq_Scalar2}) is based on small signal analysis and valid only for a diode operating in the nonlinear region. Hence, if the peak of $y_q(t)$ is so large that the diode series resistance dominates the diode behaviour and the diode I-V characteristic is linear \cite{BCCG13}, the assumptions made for deriving the nonlinear model does not hold.
Interestingly, the model (\ref{EqVoutq_Scalar2}) (derived in \cite{HC16TSParxiv, HC16SPAWC}) divided by a constant is equal to the $z_{DC}$ model (for $n_o = 4$) in \cite{CB16}, which has been validated through circuit simulations for various rectifier configurations \cite{CB16,CB17arXiv}. It means that the two models are equivalent in terms of optimization. If (\ref{EqVoutq_Scalar2}) is truncated to the 2nd-order term, i.e. only the first term (which contains $\beta_2$) in (\ref{EqVoutq_Scalar2}) is remaining, the model is referred to as a linear model\cite{CB16,ZZH13}.

\subsection{A Compact Nonlinear Model}\label{SecCompactExpress}
\begin{figure}[!t]
\centering
\includegraphics[width = 3.4in]{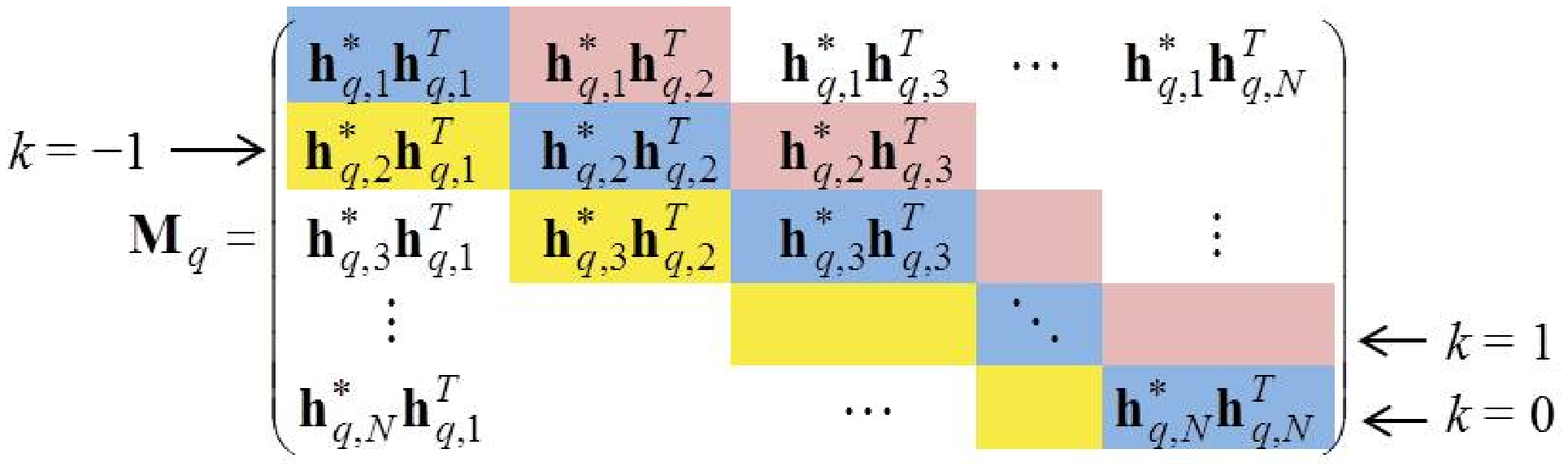}
\caption{$\mathbf{M}_{q,1}$ is the above matrix only maintaining the block diagonal (whose index is $k=1$) in pink, while all the other blocks are set as $\mathbf{0}_{M\times M}$.}
\label{FigM_mat}
\end{figure}
In order to formulate a compact model, $s_{(n-1)M+m}$ is collected into a vector $\mathbf{s} = [\mathbf{s}_1^T, \ldots, \mathbf{s}_N^T]^T \!\in\! \mathbb{C}^{MN \times 1}$, where $\mathbf{s}_n = [s_{(n-1)M+1}, \ldots, s_{(n-1)M+M}]^T$ collects precoding weights of the signals transmitted by the $M$ antennas at frequency $n$. In the following sections, $\mathbf{s}$ is referred to as a waveform precoder, which collects magnitudes and phases (in space and frequency domains) for generating multi-sine signals.

Channel gain $h_{q,(n-1)M + m}$ is collected into $\mathbf{h}_q = [\mathbf{h}_{q,1}^T, \ldots,$ $\mathbf{h}_{q,N}^T]^T \! \in \! \mathbb{C}^{MN \times 1}$, where $\mathbf{h}_q$ collects the spatial/frequency domain channel gains with respect to (w.r.t.) rectenna $q$, and $\mathbf{h}_{q,n} = [h_{q,(n-1)M+1}, \ldots, h_{q,(n-1)M+M}]^T$ characterizes the spatial domain channel between the ET and rectenna $q$ at frequency $n$. By substituting (\ref{EqYq_tilde}) and (\ref{EqYq}) into (\ref{EqVoutq_Scalar2}), $v_{\text{out},q}$ can be expressed as a function of the precoder $\mathbf{s}$. As $f_\text{LPF}(\cdot)$ removes all the non-DC component, (\ref{EqVoutq_Scalar2}) can be finally reformulated as
\begin{IEEEeqnarray}{l}
\label{Eqv_outInhomo}
v_{\text{out},q} = \beta_2 \textstyle{\sum_{n = 1}^N} \mathbf{s}_n^H \mathbf{h}_{q,n}^\ast \mathbf{h}_{q,n}^T \mathbf{s}_n + \nonumber\\
\frac{3}{2} \beta_4 \! \sum_{\substack{{n_1, n_2, n_3, n_4}\\{n_1 \!-\!  n_3\!  =\!  -\!  (n_2\!  -\!  n_4)}}} \! \mathbf{s}_{n_3}^H \! \mathbf{h}_{q,n_3}^\ast \! \mathbf{h}_{q,n_1}^T \! \mathbf{s}_{n_1} \!\cdot\! \mathbf{s}_{n_4}^H \! \mathbf{h}_{q,n_4}^\ast \! \mathbf{h}_{q,n_2}^T \! \mathbf{s}_{n_2}.
\end{IEEEeqnarray}
The above (\ref{Eqv_outInhomo}) can be transformed into a more compact form, by introducing {\small$MN$}-by-{\small$MN$} matrices
\begin{equation}
\label{EqMatM_q}
\mathbf{M}_q \triangleq \mathbf{h}_q^\ast \mathbf{h}_q^T
\end{equation}
and $\mathbf{M}_{q,k}$. As shown in Fig. \ref{FigM_mat}, $k \! \in \! \{1, \ldots, N\!-\!1\}$ is the index of the $k$\,th block diagonal above the main block diagonal (whose index $k=0$) of $\mathbf{M}_q$, while $k \! \in \! \{-\!(N\!-\!1), \ldots, -1\}$ is the index of the $|k|$\,th block diagonal below the main block diagonal. Given a certain $k$, $\mathbf{M}_{q,k}$ is generated by retaining the $k$\,th block diagonal of $\mathbf{M}_q$ but setting all the other blocks as $\mathbf{0}_{M\times M}$. For $k\neq 0$, the non-Hermitian matrix $\mathbf{M}_{q,-k} = \mathbf{M}_{q,k}^H$, while $\mathbf{M}_{q,0} \succeq 0$. As the matrix $\mathbf{M}_q$ defined in (\ref{EqMatM_q}) is essentially a matrix function of random channel gains $\mathbf{h}_q$, the rectenna DC output voltage $v_{\text{out},q}$ is essentially a function of $\mathbf{h}_q$ and the waveform precoder $\mathbf{s}$. Therefore,
\begin{IEEEeqnarray}{rcl}
\label{EqFuncVoutq}
v_{\text{out},q}\left(\mathbf{h}_q, \mathbf{s}\right) & {}={} & \beta_2 \mathbf{s}^H \mathbf{M}_{q,0} \mathbf{s} \! + \! \frac{3}{2}\beta_4 \mathbf{s}^H \mathbf{M}_{q,0} \mathbf{s}\!\left(\mathbf{s}^H \mathbf{M}_{q,0}\mathbf{s}\right)^H \! + \nonumber\\
&&\! 3\beta_4 \sum_{k = 1}^{N-1} \mathbf{s}^H \mathbf{M}_{q,k} \mathbf{s} \!\left(\mathbf{s}^H \mathbf{M}_{q,k} \mathbf{s}\right)^H\,.
\end{IEEEeqnarray}
\emph{Remark:} The second term in (\ref{Eqv_outInhomo}) can be written as $\frac{3}{2} \beta_4 \sum_{\substack{{n_1, n_2, n_3, n_4}\\{n_1 = n_3 \text{ and } n_2 = n_4}}} \mathbf{s}_{n_3}^H \! \mathbf{h}_{q,n_3}^\ast \mathbf{h}_{q,n_1}^T  \mathbf{s}_{n_1} \mathbf{s}_{n_4}^H  \mathbf{h}_{q,n_4}^\ast  \mathbf{h}_{q,n_2}^T  \mathbf{s}_{n_2} + \frac{3}{2} \beta_4  \sum_{\substack{{n_1, n_2, n_3, n_4}\\{n_1 -  n_3  =  -  (n_2  -  n_4) > 0}}} \mathbf{s}_{n_3}^H \! \mathbf{h}_{q,n_3}^\ast \mathbf{h}_{q,n_1}^T \! \mathbf{s}_{n_1} \mathbf{s}_{n_4}^H \! \mathbf{h}_{q,n_4}^\ast \! \mathbf{h}_{q,n_2}^T \! \mathbf{s}_{n_2} + \frac{3}{2} \beta_4  \sum_{\substack{{n_1, n_2, n_3, n_4}\\{n_1 -  n_3  =  -  (n_2  -  n_4) < 0}}}  \mathbf{s}_{n_3}^H  \mathbf{h}_{q,n_3}^\ast  \mathbf{h}_{q,n_1}^T \mathbf{s}_{n_1} \mathbf{s}_{n_4}^H  \mathbf{h}_{q,n_4}^\ast \mathbf{h}_{q,n_2}^T  \mathbf{s}_{n_2}$. Remarkably, $\mathbf{h}_{q,n_3}^\ast \mathbf{h}_{q,n_1}^T$ is within the $(n_1\!-\!n_3)$\,th block diagonal of $\mathbf{M}_q$. Hence, by defining $k \! \triangleq \! n_1 \! - \! n_3$ (therefore, $n_2 - n_4 = -k$) and applying the property $\mathbf{M}_{q,-k} = \mathbf{M}_{q,k}^H$, we can find that the second term in (\ref{Eqv_outInhomo}) can be written as the sum of the second and the third terms in (\ref{EqFuncVoutq}).

Note that in (\ref{EqFuncVoutq}), values of $\mathbf{s}^H \mathbf{M}_{q,k} \mathbf{s}$ for $k\neq 0$ are complex. Supposing that the RF transmit power at the ET is subject to $P$, $\mathbf{s}$ is subject to $\|\mathbf{s}\|^2 = \sum_{n = 1}^N \sum_{m = 1}^M |s_{(n-1)M + m}|^2 \leq P$.

\subsection{Channel Realizations for Codebook Designs}\label{SecImperfectCSIT}
This paper relaxes the assumption of perfect instantaneous CSIT. The studied multi-antenna multi-sine WPT system generates WPT waveforms from codebooks of waveform precoders, relying on limited feedback.
Hence, we assume that designing codebooks offline can exploit a sample of $T_0$ channel realizations, which can be obtained by empirical observations or the Monte Carlo sampling method \cite{SDR09} (provided statistical CSI is known).
By defining a random matrix variable $\mathbf{H} \in \mathbb{C}^{MN \times Q}$ and collecting $\mathbf{h}_q$ into the $q$\,th column of $\mathbf{H}$, i.e. $[\mathbf{H}]_q = \mathbf{h}_q$, the sample can be defined as $\mathcal{H}_0 \triangleq \{\mathbf{H}^{[1]}, \ldots, \mathbf{H}^{[T_0]} \}$, where each $\mathbf{H}^{[t_0]}$ characterizes a realization of the channel between the ET and the $Q$ rectennas at the ER across all the $N$ frequencies. Additionally, we use a vector variable $\mathbf{h}_q^{[t_0]} \triangleq \big[\mathbf{H}^{[t_0]}\big]_q$ to characterize the $t_0$\,th realization of the wireless channel between the ET and rectenna $q$ at all the $N$ frequencies. The sample $\mathcal{H}_0$ would be utilized to design the codebooks for the precoding schemes proposed in Sections \ref{SecBeamSelectProtocol} and \ref{SecBeamTrainProtocol}.

\section{Sample Average Approximation Algorithm}\label{SecMaxEwsumVout}
As a preliminary to Sections \ref{SecBeamSelectProtocol} and \ref{SecBeamTrainProtocol}, this section tackles the issue of solving an SAA problem, which approximates the problem of maximizing the expected $v_\text{out}$. The proposed SAA algorithm lays the basis for deriving codebooks based on the sample $\mathcal{H}_0$.

\subsection{Problem Formulation}
\label{SecProbFormMaxEwsumVout}
For generality, we investigate the weighted sum $v_\text{out}$. Defining $w_q \geq 0$ as the weight for rectenna $q$, the expected value of the weighted sum $v_\text{out}$ can be expressed as $\mathcal{E}\{\sum_{q=1}^Q w_q v_{\text{out},q}\left(\mathbf{h}_q, \mathbf{s}\right)\}$, where the channel gain vector $\mathbf{h}_{q}$ is a random variable, while $\mathbf{s}$ is a deterministic variable subject to a power constraint. In order to maximize the expected weighted sum $v_\text{out}$, a stochastic programming problem can be formulated as
\begin{IEEEeqnarray}{cl}
\label{Prob_StochProMaxVout}
\max_{\mathbf{s}} \quad & \mathcal{E}\left\{ \textstyle{\sum_{q=1}^Q} w_q v_{\text{out},q}\left(\mathbf{h}_q, \mathbf{s}\right)\right\} \IEEEyesnumber \IEEEyessubnumber \label{EqEWSumVout}\\
\text{s.t.} \,& \|\mathbf{s}\|^2 \leq P\,.\IEEEyessubnumber
\end{IEEEeqnarray}
To solve (\ref{Prob_StochProMaxVout}) and lay the foundations for deriving a codebook based on a sample, we formulate a SAA problem \cite{SDR09} of (\ref{Prob_StochProMaxVout}) based on the sample $\mathcal{H}_0$ and approximate the optimal $\mathbf{s}^\star$ of (\ref{Prob_StochProMaxVout}). Under the assumption of $\mathcal{H}_0$, the expected value in (\ref{EqEWSumVout}) can be approximated by the average value $\sum_{q = 1}^Q \frac{1}{T_0} \sum_{t_0 = 1}^{T_0} w_q v_{\text{out},q}(\mathbf{h}_q^{[t_0]}, \mathbf{s})$. The SAA problem of (\ref{Prob_StochProMaxVout}) can be formulated as
\begin{IEEEeqnarray}{cl}
\label{Prob_MaxVout_SAA}
\max_{\mathbf{s}} \quad & \sum_{t_0 = 1}^{T_0} \sum_{q=1}^Q w_q v_{\text{out},q}\left(\mathbf{h}_q^{[t_0]}, \mathbf{s}\right) \IEEEyesnumber \IEEEyessubnumber \label{Eq_AvgVout_MaxVout_SAA}\\
\text{s.t.} \,& \|\mathbf{s}\|^2 \leq P\,, \IEEEyessubnumber
\end{IEEEeqnarray}
where
\begin{IEEEeqnarray}{rcl}
\label{Eq_AvgVout_MaxVout_SAA_Expan}
v_{\text{out},q}\!\left(\!\mathbf{h}_q^{[t_0]}, \mathbf{s}\!\right) & = & \beta_2 \mathbf{s}^H \mathbf{M}_{q,0}^{[t_0]} \mathbf{s} \! + \! \frac{3}{2}\beta_4 \mathbf{s}^H \mathbf{M}_{q,0}^{[t_0]} \mathbf{s}\!\left(\!\mathbf{s}^H \mathbf{M}_{q,0}^{[t_0]} \mathbf{s}\!\right)^{\!H} \! + \nonumber\\
&&\! 3\beta_4 \sum_{k = 1}^{N-1} \mathbf{s}^H \mathbf{M}_{q,k}^{[t_0]} \mathbf{s} \!\left(\!\mathbf{s}^H \mathbf{M}_{q,k}^{[t_0]} \mathbf{s}\!\right)^H\,.
\end{IEEEeqnarray}
and the matrix $\mathbf{M}_{q,k}^{[t_0]}$ is obtained by $\mathbf{M}_q^{[t_0]} \triangleq \big[\mathbf{h}_q^{[t_0]}\big]^\ast \big[\mathbf{h}_q^{[t_0]}\big]^T$. The constant $1/T_0$ has been eliminated from (\ref{Eq_AvgVout_MaxVout_SAA}), without effect on the optimal $\mathbf{s}^\star$. It can be shown by the Law of Large Numbers that (\ref{Eq_AvgVout_MaxVout_SAA}) multiplied by $1/T_0$ converges to (\ref{EqEWSumVout}), as $T_0$ tends to infinity\cite{SDR09}.

\emph{Remark:} In the GLA for MIMO communications\cite{XG06,CO13}, an SAA problem boils down to maximizing the average signal-to-noise ratio. To solve this SAA problem, a channel correlation matrix (i.e. a second-moment matrix) is first approximated by exploiting a sample of channel realizations. Then, an optimum turns out to be the dominant eigenvector of the correlation matrix. Nevertheless, due to the structure of the 4\,th-order term in the nonlinear model (\ref{Eqv_outInhomo}), we cannot solve (\ref{Prob_MaxVout_SAA}) by approximating a second-moment matrix or a fourth-moment matrix.

Problem (\ref{Prob_MaxVout_SAA}) is intractable, due to the quartic polynomial (\ref{Eq_AvgVout_MaxVout_SAA_Expan}). To tackle this issue, we make use of the optimization framework developed in \cite{HC16TSParxiv, HC16SPAWC}. Specifically, auxiliary variables $t_{q,k}^{[t_0]}$ and $\mathbf{X}$ are introduced to linearize $\mathbf{s}^H \mathbf{M}_{q,k}^{[t_0]} \mathbf{s}$ $\forall k \! \in \! \{0,\ldots,N-1\}$ , such that
\begin{equation}
\mathbf{s}^H \mathbf{M}_{q,k}^{[t_0]} \mathbf{s} = \text{Tr}\left\{ \mathbf{M}_{q,k}^{[t_0]} \mathbf{s}\mathbf{s}^H\right\} = \text{Tr}\left\{ \mathbf{M}_{q,k}^{[t_0]} \mathbf{X}\right\} = t_{q,k}^{[t_0]}\,.
\end{equation}
Hence, problem (\ref{Prob_MaxVout_SAA}) can be recast into an equivalent form
\begin{IEEEeqnarray}{cl}
\label{Epi_Prob_MaxVoutSAA}
\min_{\gamma_1, t_{q,k}^{[t_0]}, \mathbf{X}\succeq 0} \, & \gamma_1 \IEEEyesnumber\IEEEyessubnumber\\
\text{s.t.} & \sum_{t_0\!=\!1}^{T_0} \sum_{q\!=\!1}^Q w_q \! \left(\!- \beta_2 t_{q,0}^{[t_0]}\! +\! g_q^{[t_0]} \! \left(\! \mathbf{t}_q^{[t_0]} \!\right)\! \right) \! - \! \gamma_1 \! \leq \! 0, \IEEEyessubnumber \label{Eq_MaxVoutSAA_NCVXqc}\\
& \text{Tr}\Big\{ \mathbf{M}_{q,k}^{[t_0]} \mathbf{X}\Big\} = t_{q,k}^{[t_0]}\,,\, \forall t_0, q, k\,, \IEEEyessubnumber \label{EqConst_t_t0qk}\\
& \text{Tr}\Big\{\! \left[\mathbf{M}_{q,k}^{[t_0]}\right]^H \mathbf{X} \! \Big\} = \left[t_{q,k}^{[t_0]}\right]^\ast,\forall t_0, q, k\neq0,\quad \IEEEyessubnumber \label{EqConstCjgt_t0qk}\\
& \text{Tr}\{\mathbf{X}\} \leq P\,, \IEEEyessubnumber \label{EqTxPwrConstSAA}\\
& \text{rank}\{\mathbf{X}\} = 1\,, \IEEEyessubnumber \label{EqEpiMaxVoutSAARankConst}
\end{IEEEeqnarray}
where $\mathbf{A}_0 =  diag\{-3\beta_4/2, -3\beta_4, \ldots, -3\beta_4\}  \preceq  0$,
\begin{equation}
\mathbf{t}_q^{[t_0]} \! \triangleq \! \left[\! t_{q,0}^{[t_0]},\ldots,t_{q,N-1}^{[t_0]} \!\right]^T \text{and } g_q^{[t_0]}\left(\mathbf{t}_q^{[t_0]}\right) \! \triangleq \! \left[\mathbf{t}_q^{[t_0]}\right]^H \! \mathbf{A}_0  \mathbf{t}_q^{[t_0]}\,.
\end{equation}
Although the quartic polynomial problem has been addressed, problem (\ref{Epi_Prob_MaxVoutSAA}) is still NP-hard in general, due to the nonconvex quadratic constraint (\ref{Eq_MaxVoutSAA_NCVXqc}) and the rank constraint (\ref{EqEpiMaxVoutSAARankConst}).

\subsection{Sample Average Approximation Algorithm}
This subsection proposes an algorithm to solve the equivalent form of (\ref{Prob_MaxVout_SAA}), i.e. problem (\ref{Epi_Prob_MaxVoutSAA}). We first relax the rank constraint in problem
\begin{equation}
\label{Epi_Prob_MaxVoutSAA_RkRelxd}
\min_{\gamma_1, t_{q,k}^{[t_0]}, \mathbf{X}\succeq 0} \{\gamma_1: \text{(\ref{Eq_MaxVoutSAA_NCVXqc}), (\ref{EqConst_t_t0qk}), (\ref{EqConstCjgt_t0qk}) and (\ref{EqTxPwrConstSAA})} \}\,.
\end{equation}
In the following, we will show that solving (\ref{Epi_Prob_MaxVoutSAA_RkRelxd}) can yield a rank-1 optimized $\mathbf{X}^\star$, such that this $\mathbf{X}^\star$ is a solution of problem (\ref{Epi_Prob_MaxVoutSAA}).

As (\ref{Eq_MaxVoutSAA_NCVXqc}) is a nonconvex quadratic constraint, we exploit Successive Convex Approximation (SCA) to address (\ref{Epi_Prob_MaxVoutSAA_RkRelxd}). Specifically, given $\big[\mathbf{t}_q^{[t_0]}\big]^{(l-1)}$ as the optimal $\mathbf{t}_q^{[t_0]}$ achieved at iteration $(l-1)$, the nonconvex term $g_q^{[t_0]}\big(\mathbf{t}_q^{[t_0]}\big)$ in (\ref{Eq_MaxVoutSAA_NCVXqc}) can be approximated as an affine function
\begin{IEEEeqnarray}{l}
\tilde{g}_q^{[t_0]}\left(\mathbf{t}_q^{[t_0]}, \big[\mathbf{t}_q^{[t_0]}\big]^{(l-1)}\right) \triangleq 2\text{Re} \left\{ \left[ \big[\mathbf{t}_q^{[t_0]}\big]^{(l-1)} \right]^H \! \mathbf{A}_0  \mathbf{t}_q^{[t_0]} \right\} \nonumber\\
\quad\quad\quad\quad\quad\quad\quad\quad - \left[ \big[\mathbf{t}_q^{[t_0]}\big]^{(l-1)} \right]^H \mathbf{A}_0 \big[\mathbf{t}_q^{[t_0]}\big]^{(l-1)}
\end{IEEEeqnarray}
at iteration $l$, by the first-order Taylor expansion. The approximate problem of (\ref{Epi_Prob_MaxVoutSAA_RkRelxd}) at iteration $l$ can be formulated as
\begin{IEEEeqnarray}{l}
\label{Epi_Prob_MaxVoutSAA_AP}
\min_{\gamma_1, t_{q,k}^{[t_0]}, \mathbf{X}\succeq 0} \, \gamma_1 \IEEEyesnumber\IEEEyessubnumber\\
\text{s.t.}\, \sum_{t_0 = 1}^{T_0} \! \sum_{q = 1}^Q w_q \! \left(\! - \beta_2 t_{q,0}^{[t_0]} \!+\! \tilde{g}_q^{[t_0]}\! \left(\mathbf{t}_q^{[t_0]}\!, \big[\mathbf{t}_q^{[t_0]}\big]^{(l - 1)}  \right)  \! \right) \! \leq \!  \gamma_1,\quad\quad\IEEEyessubnumber \label{Eq_MaxVoutSAA_NCVXqcAP}\\
\quad \, \text{(\ref{EqConst_t_t0qk}), (\ref{EqConstCjgt_t0qk}) and (\ref{EqTxPwrConstSAA})}\,.\nonumber
\end{IEEEeqnarray}
The above approximate problem is convex. By substituting (\ref{EqConst_t_t0qk}) and (\ref{EqConstCjgt_t0qk}) into (\ref{Eq_MaxVoutSAA_NCVXqcAP}), the above problem can be recast into an equivalent but more concise form
\begin{IEEEeqnarray}{cl}
\label{MaxVoutSAA_AP_Equiv}
\min_{\mathbf{X}\succeq 0} \, & \left\{\text{Tr}\{\mathbf{A}_1\mathbf{X}\}: \text{Tr}\{\mathbf{X}\} \leq P \right\}\,,
\end{IEEEeqnarray}
where $\mathbf{A}_1 \triangleq \mathbf{C}_1 + \mathbf{C}_1^H$ and
\begin{IEEEeqnarray}{rcl}
\label{EqC_1}
\mathbf{C}_1 & {} \triangleq {} & \sum_{t_0=1}^{T_0} \sum_{q=1}^Q w_q \left( -\frac{\beta_2 +  3 \beta_4 \big[ t_{q,0}^{[t_0]} \big]^{(l - 1)} }{2}\mathbf{M}_{q,0}^{[t_0]} \right. \nonumber \\
&& \left. {}-{} 3 \beta_4 \sum_{k=1}^{N - 1} \left[ \big[ t_{q,k}^{[t_0]} \big]^{(l - 1)} \right]^\ast \mathbf{M}_{q,k}^{[t_0]} \right)\,.
\end{IEEEeqnarray}
Problem (\ref{MaxVoutSAA_AP_Equiv}) is shown to be a separable semidefinite program (SDP)\cite{HP10}. Thus, applying \cite[Proposition 3.5]{HP10}, we can show that problem (\ref{MaxVoutSAA_AP_Equiv}) has, among others, a rank-1 optimal solution $\mathbf{X}^\star$. Due to the equivalence, the optimal $\mathbf{X}^\star$ of problem (\ref{MaxVoutSAA_AP_Equiv}) also satisfies the Karush-Kuhn-Tucker (KKT) conditions of problem (\ref{Epi_Prob_MaxVoutSAA_AP}). As problem (\ref{Epi_Prob_MaxVoutSAA_AP}) is convex, the rank-1 solution $\mathbf{X}^\star$ of problem (\ref{MaxVoutSAA_AP_Equiv}) is also the optimal $\mathbf{X}^\star$ of problem (\ref{Epi_Prob_MaxVoutSAA_AP}).

In order to achieve such a rank-1 $\mathbf{X}^\star$ for problem (\ref{MaxVoutSAA_AP_Equiv}), we can employ the Interior-Point Algorithm (IPA) to the SDP problem (\ref{MaxVoutSAA_AP_Equiv}) yielding a high-rank $\mathbf{X}^\star$\cite{GB14} , from which we can derive a rank-1 $\mathbf{X}^\star$ by rank reduction \cite{HP10} or a randomized algorithm \cite{HP14}. However, solving (\ref{MaxVoutSAA_AP_Equiv}) by IPA results in high complexity: $O(1)(2+2MN)^{1/2}(MN)^2\big(5(MN)^4 + 8(MN)^3 + (MN)^2 +1\big)$\cite{BN01}. In the following, we show that the rank-1 $\mathbf{X}^\star$ can be computed by a less complex method.

As problem (\ref{MaxVoutSAA_AP_Equiv}) has a rank-1 solution $\mathbf{X}^\star$, there always exists a vector $\mathbf{x}^\star$ such that $\mathbf{X}^\star = \mathbf{x}^\star [\mathbf{x}^\star]^H$. Therefore, computing an optimal rank-1 $\mathbf{X}^\star$ of problem (\ref{MaxVoutSAA_AP_Equiv}) boils down to looking for an optimal vector variable $\mathbf{x}^\star$ of
\begin{IEEEeqnarray}{cl}
\label{MaxVoutSAA_AP_EquivQP}
\min_{\mathbf{x}} \, & \left\{\mathbf{x}^H \mathbf{A}_1\mathbf{x}: \|\mathbf{x}\|^2 \leq P \right\}\,.
\end{IEEEeqnarray}
Intuitively, if the smallest eigenvalue $\mathbf{A}_1$ is positive, $\mathbf{x}^\star = \mathbf{0}$.
\begin{proposition}
\label{PropMatA1eigLessZero}
The matrix $\mathbf{A}_1$ in problem (\ref{MaxVoutSAA_AP_EquivQP}) always has at least one negative eigenvalue.
\end{proposition}
\begin{IEEEproof}
In (\ref{EqC_1}), diagonal entries of $\mathbf{M}_{q,0}^{[t_0]}$ are always positive, while those of $\mathbf{M}_{q,k}^{[t_0]}$ and $\big[ \mathbf{M}_{q,k}^{[t_0]} \big]^H$ (for $k \neq 0$) are 0. Hence, the diagonal entries of $\mathbf{A}_1$ are always negative and $\text{Tr}\{\mathbf{A}_1\} < 0$, from which the proposition can be shown.
\end{IEEEproof}

Given Proposition \ref{PropMatA1eigLessZero}, the KKT conditions of problem (\ref{MaxVoutSAA_AP_EquivQP}) indicate that the stationary points of problem (\ref{MaxVoutSAA_AP_EquivQP}) are in the directions of the eigenvectors of $\mathbf{A}_1$. Thus, to minimize $\mathbf{x}^H \mathbf{A}_1\mathbf{x}$, the direction of the optimal $\mathbf{x}^\star$ should be aligned with the eigenvector corresponding to the minimum eigenvalue. Hence, the optimal $\mathbf{x}^\star = \sqrt{P}[\mathbf{U}_{\mathbf{A_1}}]_{\text{min}}$, where $\mathbf{U}_{\mathbf{A_1}}$ collects the eigenvectors of $\mathbf{A}_1$, and $[\mathbf{U}_{\mathbf{A_1}}]_{\text{min}}$ represents the eigenvector corresponding to the smallest eigenvalue of $\mathbf{A}_1$.

\begin{algorithm}
\caption{SAA Algorithm ${\bf\Psi} \left(\mathcal{H}_0, \mathbf{s}_\text{int}\right)$}\label{AlgSAA}
\begin{algorithmic}[1]
\Statex \textbf{Input:}  $\mathcal{H}_0 = \{\mathbf{H}^{[1]}, \ldots, \mathbf{H}^{[T_0]} \}$ and $\mathbf{s}_\text{int}$.
\Statex \textbf{Output:} The optimized solution $\mathbf{s}^\star$ of problem (\ref{Prob_MaxVout_SAA}).
\State \textbf{Initialization} $t_0 = 1, \ldots, \text{Card}(\mathcal{H}_0)$; generate $\mathbf{M}_q^{[t_0]} = \big[\mathbf{h}_q^{[t_0]}\big]^\ast \big[\mathbf{h}_q^{[t_0]}\big]^T$ $\forall q,t_0$; set $l = 0$; initialize $\mathbf{s}^\star$ as $\mathbf{s}_\text{int}$ such that $\mathbf{X}^{(0)} = \mathbf{s}_\text{int} \mathbf{s}_\text{int}^H$ satisfies (\ref{EqTxPwrConstSAA}); compute $\big[t_{q,k}^{[t_0]}\big]^{(0)}$ $\forall t_0, q, k$ with $\mathbf{X}^{(0)}$ by (\ref{EqConst_t_t0qk});
\Repeat
    \State $l = l + 1$;
    \State Compute $\mathbf{C}_1$ by (\ref{EqC_1}), and then update $\mathbf{A}_1$;
    \State Compute the eigenvalue decomposition of $\mathbf{A}_1$ and the optimal solution of problem (\ref{MaxVoutSAA_AP_EquivQP}), i.e. $\mathbf{x}^\star = \sqrt{P}\left[\mathbf{U}_{\mathbf{A}_1}\right]_{\text{min}}$;
    \State Update the optimal $\mathbf{X}^\star$ at iteration $l$ for problems (\ref{MaxVoutSAA_AP_Equiv}) and (\ref{Epi_Prob_MaxVoutSAA_AP}), i.e. $\mathbf{X}^{(l)} = \mathbf{x}^\star[\mathbf{x}^\star]^H$;
    \State Update $\big[t_{q,k}^{[t_0]}\big]^{(l)}$ $\forall t_0, q, k$ by (\ref{EqConst_t_t0qk});
\Until{\|\mathbf{X}^{(l)} - \mathbf{X}^{(l-1)}\|_F/\|\mathbf{X}^{(l)}\|_F \leq \epsilon}
\State $\mathbf{s}^\star = \mathbf{x}^\star$. \label{AlgSCALine7}
\end{algorithmic}
\end{algorithm}
The proposed SAA algorithm for solving problem (\ref{Prob_MaxVout_SAA}) is summarized in Algorithm \ref{AlgSAA}. In each iteration, performing eigenvalue decomposition for $\mathbf{A}_1$ by the QR algorithm yields complexity of $O\big((MN)^3\big)$\cite{Parlett00}.
It is remarkable that this computational complexity and that of exploiting the IPA to solve the SDP (\ref{MaxVoutSAA_AP_Equiv}) are not a function of $T_0$ or $Q$, while the former complexity is much lower than the complexity of the IPA used to solve the SDP.

\begin{theorem}
\label{TheoAlgSAAconverge}
Algorithm \ref{AlgSAA} converges to a limit point, and $\gamma_1$ in problem (\ref{Epi_Prob_MaxVoutSAA_AP}) decreases over iterations. Designating this limit point as $\mathbf{\bar{s}}$, $\mathbf{\bar{s}} \mathbf{\bar{s}}^H$ satisfies the KKT conditions of problem (\ref{Epi_Prob_MaxVoutSAA}).
\end{theorem}
\begin{IEEEproof}
For details, see Appendix \ref{AppTheoAlgSAAconverge}.
\end{IEEEproof}
Algorithm \ref{AlgSAA} can be expressed as a function
\begin{equation}
\label{EqAlgSAA_Func}
\mathbf{s}^\star = {\bf\Psi} \left(\mathcal{H}_0, \mathbf{s}_\text{int}\right)\,.
\end{equation}
which expresses the optimized $\mathbf{s}^\star$ of problem (\ref{Prob_MaxVout_SAA}) as a function of the sample $\mathcal{H}_0$ and the initial point $\mathbf{s}_\text{int}$. In the following, ``Algorithm \ref{AlgSAA}" and the expression in (\ref{EqAlgSAA_Func}) are used interchangeably to refer to the SAA algorithm.

\section{Waveform Selection-Based WPT}\label{SecBeamSelectProtocol}
\subsection{Waveform Selection Procedure}
This subsection elaborates on the WS procedure. As shown in Fig. \ref{FigBeamSelectionProtocol}, each time frame of the WS-based WPT consists of a WS phase and a WPT phase.
The WS phase is composed of $N_p$ ET transmission timeslots and an ER feedback timeslot. During each ET transmission timeslot, the ET exploits a precoder within a predesigned $N_p$-codeword codebook to generate a multi-sine signal and send this signal. We designate the codebook as a set $\mathcal{S}$, where the $n_p$\,th codeword is expressed as a column vector and designated as $[\mathcal{S}]_{n_p}$ for $n_p = 1,\ldots,N_p$. Once the multi-sine signal is received at the ER, the ER measures the weighted sum $v_\text{out}$ for the $Q$ rectennas and records its value. After the $N_p$ ET transmission timeslots, the ER finds the index corresponding to the largest weighted sum $v_\text{out}$ value. This index is also the index $n_p^\star$ of the waveform that maximizes the weighted sum $v_\text{out}$. By defining the rectenna DC output voltage measured at the ER as $v_{\text{out},q}([\mathcal{S}]_{n_p})$, the search of the index $n_p^\star$ can be expressed as
\begin{equation}
\label{EqCdwdSelect}
n_p^\star = \arg\max_{n_p \in \{1,\ldots,N_p\}} \textstyle{\sum_{q=1}^Q} w_q \cdot v_{\text{out},q}([\mathcal{S}]_{n_p})\,.
\end{equation}
Then, at the ER feedback timeslot, the ER feeds back the index $n_p^\star$, which is composed of $\log_2 N_p$ feedback bits. The above WS procedure essentially conducts an exhaustive search over the codebook $\mathcal{S}$ and finds the codeword that maximizes the $v_\text{out}$ during the WPT phase. In the following WPT phase, the ET generates multi-sine signals for WPT by using the precoder $[\mathcal{S}]_{n_p^\star}$.

In the presence of $N_p = 1$, the WS phase is eliminated, and the ET exploits the only codeword to generate the WPT waveform. Such a scheme (where $N_p = 1$) is referred to as a special case of the WS-based WPT.

\subsection{Codebook Design for the WS-based WPT}\label{SecCdbkDesignAlgoVQ}
Compared to an ideal precoder achieved under the assumption of perfect instantaneous CSIT, the precoder $[\mathcal{S}]_{n_p^\star}$ taken from a codebook causes performance degradation in terms of the weighted sum $v_\text{out}$ (in the WPT phase). Hence, this subsection proposes an algorithm based on GLA, so as to derive a codebook $\mathcal{S}$ that minimizes the average weighted sum $v_{\text{out}}$ performance distortion.

\subsubsection{Design Problem Formulation}
The codebook $\mathcal{S}$ is derived from the sample $\mathcal{H}_0$, which is illustrated in Section \ref{SecImperfectCSIT}. To formulate the codebook design problem, we first define a distortion function, which can be exploited to quantify the average weighted sum $v_{\text{out}}$ distortion caused by the selected codeword $[\mathcal{S}]_{n_p^\star}$. Based on the model (\ref{EqFuncVoutq}), the weighted sum $v_{\text{out}}$-distortion function w.r.t. a certain channel realization $\mathbf{H}^{[t_0]}$ and the selected precoder $[\mathcal{S}]_{n_p^\star}$ can be defined as
\begin{IEEEeqnarray}{rcl}
\label{EqWsumVoutDistortFunc}
f_d \big(\mathbf{H}^{[t_0]}, \mathbf{s}_\text{opt}^{[t_0]}, [\mathcal{S}]_{n_p^\star} \big) &{}={}& \textstyle{\sum_{q=1}^Q}  w_q \! \cdot\!  v_{\text{out},q} \big(\mathbf{h}_q^{[t_0]}, \mathbf{s}_\text{opt}^{[t_0]} \big) \nonumber\\
&& {}-{} \textstyle{\sum_{q=1}^Q}  w_q \!\cdot\!  v_{\text{out},q} \big(\mathbf{h}_q^{[t_0]}, [\mathcal{S}]_{n_p^\star}\big)\!,\quad
\end{IEEEeqnarray}
where $\mathbf{s}_\text{opt}^{[t_0]}$ represents the ideal precoder computed by \cite[Algorithm 2]{HC16TSParxiv} accounting for perfect CSIT. In other words, $\mathbf{s}_\text{opt}^{[t_0]}$ is computed when $\mathbf{H}^{[t_0]}$ is perfectly known at the ET. We define an ideal precoder set $\mathcal{S}_{\text{opt}}$ to collect $\mathbf{s}_\text{opt}^{[t_0]}$, namely
\begin{equation}
\label{EqSetOptBFomer}
\mathcal{S}_{\text{opt}} = \big\{ \mathbf{s}_\text{opt}^{[1]}, \ldots, \mathbf{s}_\text{opt}^{[T_0] }\big\}\,.
\end{equation}

Given a codebook $\mathcal{S}$ consisting of $N_p$ codewords, performing the WS procedure (\ref{EqCdwdSelect}) for a given training set $\mathcal{H}_0$ essentially partitions $\mathcal{H}_0$ into $N_p$ subsets, which are designated as $\mathcal{H}_{0,n_p}$ for $n_p = 1,\ldots, N_p$. Given $[\mathcal{S}]_{n_p}$, the $n_p$\,th partition cell $\mathcal{H}_{0,n_p}$ can be obtained by
\begin{IEEEeqnarray}{rl}
\label{EqVQChannelsubset}
\mathcal{H}_{0,n_p} = & \Big\{ \mathbf{H}: f_d\Big(\!\mathbf{H}, \mathbf{s}_\text{opt}, [\mathcal{S}]_{n_p}\!\Big) \leq f_d\Big(\!\mathbf{H}, \mathbf{s}_\text{opt}, [\mathcal{S}]_{n_p^\prime}\!\Big), \nonumber\\
& \quad \forall\, \mathbf{H} \! \in \! \mathcal{H}_0, \forall\, \mathbf{s}_\text{opt} \! \in \! \mathcal{S}_{\text{opt}}, \forall\, n_p^\prime \neq n_p \Big\}. \quad
\end{IEEEeqnarray}
As each $\mathbf{s}_\text{opt}^{[t_0]}$ is computed by \cite[Algorithm 2]{HC16TSParxiv} for a given $\mathbf{H}^{[t_0]}$, $\mathcal{S}_{\text{opt}}$ can also be partitioned into $N_p$ subsets, where each subset is designated as $\mathcal{S}_{\text{opt},n_p}$. Hence, $\text{Card}(\mathcal{H}_{0,n_p}) = \text{Card}(\mathcal{S}_{\text{opt},n_p}) \triangleq T_{n_p}$. As $\mathcal{H}_0 = \bigcup_{n_p = 1}^{N_p} \mathcal{H}_{0,n_p}$, it follows that $\sum_{n_p = 1}^{N_p} T_{n_p} = T_0$. Then, we can obtain the $v_{\text{out}}$-distortion function
\begin{IEEEeqnarray}{l}
\bar{f}_d \!\left( \{\mathcal{H}_{0,n_p}\}_{n_p=1}^{N_p}, \{\mathcal{S}_{\text{opt},n_p}\}_{n_p=1}^{N_p}, \mathcal{S} \right) =
\nonumber \\
\quad\quad\quad \sum_{n_p = 1}^{N_p}\quad \sum_{\mathbf{H} \in \mathcal{H}_{0,n_p}, \mathbf{s}_\text{opt} \in \mathcal{S}_{\text{opt},n_p} } f_d \Big( \mathbf{H}, \mathbf{s}_\text{opt}, [\mathcal{S}]_{n_p} \Big).
\end{IEEEeqnarray}
Minimizing the average $v_{\text{out}}$-distortion $\frac{1}{T_0} \bar{f}_d$ amounts to minimizing the distortion $\bar{f}_d$. Due to this, the codebook design problem can be formally formulated as
\begin{equation}
\label{Prob_VQ}
\mathcal{S}^\star \!=\! \arg \min_{\mathcal{S}} \!\left\{\! \bar{f}_d: \left\|[\mathcal{S}]_{n_p}\right\|^2 \!\leq\!  P, \,\forall\, n_p \!\right\}\,.
\end{equation}

\subsubsection{Codebook Design Algorithm}\label{SecCdbkDesignAlgoVQ_AlgDesign}
It can be seen in problem (\ref{Prob_VQ}) that the partition cells $\{\mathcal{H}_{0,n_p}\}_{n_p=1}^{N_p}$ in the distortion $\bar{f}_d$ are characterized by (\ref{EqVQChannelsubset}) with optimization variables  $[\mathcal{S}]_{n_p}$. That is, the partition cells of $\{\mathcal{H}_{0,n_p}\}_{n_p=1}^{N_p}$ depend on the optimization variable $\mathcal{S}$. These coupled optimization variables make problem (\ref{Prob_VQ}) intractable.

Intuitively, if $\mathcal{H}_{0,n_p}$ and $\mathcal{S}$ were decoupled (such that they were independent optimization variables), it might be easier to solve problem (\ref{Prob_VQ}). This is reminiscent of the GLA, where partition of a given training set and codewords are optimized alternatively. Inspired by this, we then decouple the optimizations of the partition cells $\{\mathcal{H}_{0,n_p}\}_{n_p=1}^{N_p}$ and the codewords $[\mathcal{S}]_{n_p}$ in problem (\ref{Prob_VQ}), such that the partition and the codewords can be optimized alternatively. The necessary condition for optimizing $\mathcal{S}$ in (\ref{Prob_VQ}) is that the partition optimization and the codewords optimization make the distortion $\bar{f}_d$ monotonically decrease.

{\sl Partition Optimization:} The necessary condition for optimizing $\mathcal{S}$ in (\ref{Prob_VQ}) indicates that given a codebook $\mathcal{S}$, optimized partition $\{\mathcal{H}_{0,n_p}\}_{n_p=1}^{N_p}$ should always make the distortion $\bar{f}_d$ no greater than that computed with arbitrary partition $\{\mathcal{H}_{0,n_p}^\prime\}_{n_p=1}^{N_p}$, i.e. $\bar{f}_d \big( \{\mathcal{H}_{0,n_p}\}_{n_p=1}^{N_p},$ $\{\mathcal{S}_{\text{opt},n_p}\}_{n_p=1}^{N_p}, \mathcal{S} \big) \leq \bar{f}_d \big( \{\mathcal{H}_{0,n_p}^\prime\}_{n_p=1}^{N_p}, \{\mathcal{S}_{\text{opt},n_p}^\prime\}_{n_p=1}^{N_p}, \mathcal{S} \big)$. Then, it is found that the partition of $\mathcal{H}_0$ offered by (\ref{EqVQChannelsubset}) is shown to be satisfied with the above condition for optimized partition. Hence, partition cells $\big\{\mathcal{H}_{0,n_p}^{(l)}\big\}_{n_p=1}^{N_p}$ at the $l$\,th iteration of the alternating optimization should be obtained by
\begin{IEEEeqnarray}{rl}
\label{Eq_VQ_H0np_Update}
\mathcal{H}_{0,n_p}^{(l)} = \Big\{ \mathbf{H}:& f_d\big( \mathbf{H}, \mathbf{s}_\text{opt}, [\mathcal{S}]_{n_p}^{(l-1)} \big) \leq f_d\big( \mathbf{H}, \mathbf{s}_\text{opt}, [\mathcal{S}]_{n_p^\prime}^{(l-1)} \big), \nonumber\\
& \forall\, \mathbf{H} \! \in \! \mathcal{H}_0,  \forall\, \mathbf{s}_\text{opt} \! \in \! \mathcal{S}_{\text{opt}}, \forall\, n_p^\prime \neq n_p \Big\},\, \forall n_p\,,
\end{IEEEeqnarray}
where $[\mathcal{S}]_{n_p}^{(l-1)}$ and $[\mathcal{S}]_{n_p^\prime}^{(l-1)}$ are the codewords (within the codebook $\mathcal{S}^{(l-1)}$) optimized in the previous iteration. As $\mathcal{H}_0$ is a finite set, elements belonging to $\mathcal{H}_{0,n_p}^{(l)}$ and satisfying (\ref{Eq_VQ_H0np_Update}) can be found by conducting an exhaustive search over $\mathcal{H}_0$.

{\sl Codewords Optimization:} In order to always achieve the smallest $\bar{f}_d$ for given partition cells $\{ \mathcal{H}_{0,n_p}^\prime \}_{n_p=1}^{N_p}$, we need to obtain the global optimum of
\begin{equation}
\label{Prob_VQ_OriCentroidCondition}
\min_{\mathcal{S}}\! \left\{ \! \bar{f}_d\big( \{\! \mathcal{H}_{0,n_p}^\prime\!\}_{n_p=1}^{N_p},\! \{\! \mathcal{S}_{\text{opt},n_p}^\prime \!\}_{n_p=1}^{N_p},\! \mathcal{S} \big)\!: \left\|[\mathcal{S}]_{n_p}\right\|^2 \!\leq\! P, \forall n_p \!\right\}\!.
\end{equation}
The above problem boils down to $N_p$ subproblems, each corresponding to a given partition cell. The $n_p$\,th subproblem that optimizes $[\mathcal{S}]_{n_p}$ can be reformulated as
\begin{equation}
\min_{[\mathcal{S}]_{n_p}} \left\{ \textstyle{\sum_{\substack{{\mathbf{H} \in \mathcal{H}_{0,n_p}^\prime}\\{\mathbf{s}_\text{opt} \in \mathcal{S}_{\text{opt},n_p}^\prime}}} } f_d \big( \mathbf{H}, \mathbf{s}_\text{opt}, [\mathcal{S}]_{n_p} \big): \left\|[\mathcal{S}]_{n_p}\right\|^2\leq P \right\}.
\end{equation}
Defining $T_{n_p}^\prime \triangleq \text{Card}(\mathcal{H}_{0,n_p}^\prime)$, the above problem can be further recast as
\begin{IEEEeqnarray}{l}
\label{Prob_VQ_CentroidCondition_equi}
\max_{ [\mathcal{S}]_{n_p} } \left\{\! \sum_{t_{n_p} = 1}^{T_{n_p}^\prime} \sum_{q=1}^Q w_q v_{\text{out},q}\!\left(\!\mathbf{h}_q^{[t_{n_p}]}, [\mathcal{S}]_{n_p}\!\right)\!:\! \left\|[\mathcal{S}]_{n_p}\right\|^2\!\leq\! P \!\right\}\,.
\end{IEEEeqnarray}
Problem (\ref{Prob_VQ_CentroidCondition_equi}) is in the same form as problem (\ref{Prob_MaxVout_SAA}). Therefore, we can transform (\ref{Prob_VQ_CentroidCondition_equi}) into an equivalent form in the same way as how we transform (\ref{Prob_MaxVout_SAA}) into (\ref{Epi_Prob_MaxVoutSAA}). Then, $[\mathcal{S}]_{n_p}$ can be computed by Algorithm \ref{AlgSAA}.
Intuitively, we can obtain a global optimum of (\ref{Prob_VQ_OriCentroidCondition}), such that the distortion $\bar{f}_d$ monotonically decreases over iterations of partition optimization and codewords optimization, and therefore a codebook design algorithm can converge.
Nevertheless, Theorem \ref{TheoAlgSAAconverge} implies that when Algorithm \ref{AlgSAA} converges, the obtained $[\mathcal{S}]_{n_p}$ leads to a KKT point of the aforementioned equivalent form of (\ref{Prob_VQ_CentroidCondition_equi}).
This means that the solution achieved by Algorithm \ref{AlgSAA} may not be a global optimum of (\ref{Prob_VQ_OriCentroidCondition}). Fortunately, the following Proposition \ref{PropGLA_VQcdwdOpt} reveals that obtaining a global optimum of (\ref{Prob_VQ_OriCentroidCondition}) is not the only way to keep the distortion $\bar{f}_d$ monotonically decreasing.
\begin{proposition}
\label{PropGLA_VQcdwdOpt}
Given partition $\{\mathcal{H}_{0,n_p}^\prime\}_{n_p=1}^{N_p}$ and a codebook $\mathcal{S^\prime}$, setting the initial point $\mathbf{s}_\text{int} = [\mathcal{S^\prime}]_{n_p}$ and performing Algorithm \ref{AlgSAA} (i.e. $[\mathcal{S}]_{n_p} = {\bf\Psi} \big(\mathcal{H}_{0,n_p}^\prime, [\mathcal{S^\prime}]_{n_p}\big) \, \forall n_p$) always yield an optimized codebook $\mathcal{S}$ satisfying $\bar{f}_d \big( \{\mathcal{H}_{0,n_p}^\prime\}_{n_p=1}^{N_p}, \{ \mathcal{S}_{\text{opt},n_p}^\prime \}_{n_p=1}^{N_p}, \mathcal{S} \big) \leq \bar{f}_d \big( \{\mathcal{H}_{0,n_p}^\prime\}_{n_p=1}^{N_p}, \{ \mathcal{S}_{\text{opt},n_p}^\prime \}_{n_p=1}^{N_p}, \mathcal{S}^\prime \big)$.
\end{proposition}
\begin{IEEEproof}
Please see Appendix \ref{AppPropGLA_VQcdwdOpt} for details.
\end{IEEEproof}
According to Proposition \ref{PropGLA_VQcdwdOpt}, when solving (\ref{Prob_VQ_CentroidCondition_equi}) at iteration $l$ of the alternating optimization, we set the initial point $\mathbf{s}_\text{int} = [\mathcal{S}]_{n_p}^{(l-1)}$, which is the codeword obtained at iteration $(l-1)$. Hence,
\begin{equation}
\label{Eq_VQ_Snp_Update}
[\mathcal{S}]_{n_p}^{(l)} = {\bf\Psi} \big(\mathcal{H}_{0,n_p}^{(l)}, [\mathcal{S}]_{n_p}^{(l-1)}\big).
\end{equation}

{\sl Codebook Initialization:} Before performing alternating optimization, the codebook $\mathcal{S}^{(0)} \triangleq \big\{[\mathcal{S}]_1^{(0)}, \ldots, [\mathcal{S}]_{N_p}^{(0)}\big\}$ is initialized by the pruning method\cite{GG92}, which eliminates elements from $\mathcal{S}_{\text{opt}}$ and uses the remaining as a codebook. Briefly, for $n_p = 1$, the codeword is initialized as $[\mathcal{S}]_1^{(0)} = \mathbf{s}_\text{opt}^{[1]}$. Intuitively, for a given channel realization, $[\mathcal{S}]_2^{(0)}$ and $[\mathcal{S}]_1^{(0)}$ should be sufficiently distinct such that they lead to different weighted sum $v_\text{out}$. Therefore, the $\mathbf{s}_\text{opt}^{[t_0]}$ satisfying $f_d \big(\mathbf{H}^{[t_0]}, \mathbf{s}_\text{opt}^{[t_0]},  [\mathcal{S}]_1^{(0)} \big) \geq \epsilon$ is selected as $[\mathcal{S}]_2^{(0)}$. For $n_p > 2$, the codewords are initialized in a similar way. The detailed initialization algorithm is elaborated in Algorithm \ref{AlgIntializeCdbk_VQ}.

\begin{algorithm}
\caption{Codebook Initialization}\label{AlgIntializeCdbk_VQ}
\begin{algorithmic}[1]
\Statex \textbf{Input:} $\mathcal{H}_0$, $\mathcal{S}_{\text{opt}}$ and $N_p$.
\Statex \textbf{Output:} $\mathcal{S}^{(0)} = \big\{[\mathcal{S}]_1^{(0)}, \ldots, [\mathcal{S}]_{N_p}^{(0)}\big\}$;
\State \textbf{Initialization} $n_p = 1$; $t_0 = 1$;
\State $[\mathcal{S}]_{n_p}^{(0)} = \mathbf{s}_\text{opt}^{[t_0]}$; $n_p  = n_p + 1$; $t_0  = t_0 + 1$;
\Repeat
    \State $f_{d,\min}\! =\! \min_{n_p^\prime \in \{1,\ldots,n_p-1\}}\! f_d \! \big(\!\mathbf{H}^{[t_0]},\! \mathbf{s}_\text{opt}^{[t_0]},\! [\mathcal{S}]_{n_p^\prime}\! \big)$;
    \State \textbf{if} $f_{d,\min} \geq \epsilon$
    \State \quad $[\mathcal{S}]_{n_p}^{(0)} = \mathbf{s}_\text{opt}^{[t_0]}$; $n_p  = n_p + 1$; $t_0  = t_0 + 1$;
    \State \textbf{else}
    \State \quad $t_0  = t_0 + 1$;
    \State \textbf{end}
\Until{n_p > N_p}
\end{algorithmic}
\end{algorithm}

\begin{algorithm}
\caption{{\small Codebook Design Algorithm ${\bf\Phi}\big(\mathcal{H}_0, \mathcal{S}_\text{opt}, \mathcal{S}^{(0)}, N_p \big)$}}\label{AlgVQcdbk}
\begin{algorithmic}[1]
\Statex \textbf{Input:} $\mathcal{H}_0$, $\mathcal{S}_{\text{opt}}$, $\mathcal{S}^{(0)}$ and $N_p$.
\Statex \textbf{Output:} The optimized codebook $\mathcal{S}^\star$.
\State \textbf{Initialization} Set $l = 0$;
\Repeat
    \State $l = l + 1$;
    \State Partition optimization: Update $\big\{\mathcal{H}_{0,n_p}^{(l)}\big\}_{n_p=1}^{N_p}$ by (\ref{Eq_VQ_H0np_Update}) with $\mathcal{S}^{(l-1)}$; update $\mathcal{S}_{\text{opt},n_p}^{(l)}$ corresponding to $\mathcal{H}_{0,n_p}^{(l)}$ $\forall n_p$;
    \State \textbf{if} $l = 1$
    \State \quad $\bar{f}_d^{(0)} = \bar{f}_d \big(\!\{\mathcal{H}_{0,n_p}^{(1)}\}_{n_p=1}^{N_p}, \{\mathcal{S}_{\text{opt},n_p}^{(1)}\}_{n_p=1}^{N_p}, \mathcal{S}^{(0)} \big)$;
    \State \textbf{end}
    \State Codewords optimization: Update $\big\{[\mathcal{S}]_{n_p}^{(l)}\big\}_{n_p=1}^{N_p}$ by (\ref{Eq_VQ_Snp_Update});
    \State Update $\mathcal{S}^{(l)} = \big\{[\mathcal{S}]_1^{(l)}, \ldots, [\mathcal{S}]_{N_p}^{(l)}\big\}$;
    \State Update $\bar{f}_d^{(l)}  =  \bar{f}_d \big( \{\mathcal{H}_{0,n_p}^{(l)}\}_{n_p=1}^{N_p},  \{\mathcal{S}_{\text{opt},n_p}^{(l)}\}_{n_p=1}^{N_p},$ $\mathcal{S}^{(l)} \big)$;
\Until{|\bar{f}_d^{(l)} - \bar{f}_d^{(l-1)}|/|\bar{f}_d^{(l)}| \leq \epsilon^\prime}
\State $\mathcal{S}^\star = \mathcal{S}^{(l)}$.
\end{algorithmic}
\end{algorithm}

Algorithm \ref{AlgVQcdbk} summarizes the alternating optimization algorithm for deriving the codebook $\mathcal{S}$. Thanks to the partition optimization (\ref{Eq_VQ_H0np_Update}) and the codewords optimization (\ref{Eq_VQ_Snp_Update}), the sum distortion $\bar{f}_d$ monotonically decreases over iterations, satisfying $\bar{f}_d \big( \{\mathcal{H}_{0,n_p}^{(l)}\}_{n_p=1}^{N_p}, \{\mathcal{S}_{\text{opt},n_p}^{(l)}\}_{n_p=1}^{N_p}, \mathcal{S}^{(l)} \big) \leq \bar{f}_d \big( \{\mathcal{H}_{0,n_p}^{(l)}\}_{n_p=1}^{N_p}, \{\mathcal{S}_{\text{opt},n_p}^{(l)}\}_{n_p=1}^{N_p}, \mathcal{S}^{(l-1)} \big) \!\leq\! \bar{f}_d \big( \{\mathcal{H}_{0,n_p}^{(l-1)}\}_{n_p=1}^{N_p},$ $\{\mathcal{S}_{\text{opt},n_p}^{(l-1)}\}_{n_p=1}^{N_p}, \mathcal{S}^{(l-1)} \big)$. Hence, Algorithm \ref{AlgVQcdbk} converges. Furthermore, according to Theorem \ref{TheoAlgSAAconverge}, when Algorithm \ref{AlgVQcdbk} converges, the finally optimized $[\mathcal{S}]_{n_p}^\star$ satisfies the KKT conditions of the aforementioned equivalent form of (\ref{Prob_VQ_CentroidCondition_equi}).

We also express Algorithm \ref{AlgVQcdbk} as a function
\begin{equation}
\label{EqAlgVQcdbkDesign_Func}
\mathcal{S}^\star = {\bf\Phi}\big(\mathcal{H}_0, \mathcal{S}_\text{opt}, \mathcal{S}^{(0)}, N_p \big)\,.
\end{equation}
In the following sections, when we refer to the codebook design algorithm, the function expression in (\ref{EqAlgVQcdbkDesign_Func}) and ``Algorithm \ref{AlgVQcdbk}" are used interchangeably.

\section{Waveform Refinement-Based WPT}\label{SecBeamTrainProtocol}
In the WS-based WPT, the WS boils down to an exhaustive search over a codebook, where the search complexity is proportional to $N_p$. Although a large $N_p$ can increase the codewords resolution and may improve the average $v_\text{out}$ (in the WPT phase), searching among a large number of codewords can lead to a long WS phase, i.e. a high overhead, which may lower the average $v_\text{out}$ computed for the entire time frame consisting of the WS and the WPT phases.
To avoid a potentially high overhead caused by the WS procedure, this section proposes the WR-based WPT, whose time frame consists of a WR phase and a WPT phase, as shown in Fig. \ref{FigBeamTrainProtocol_ColocUsers}. Compared to the multi-bit feedback in the WS, it is easier for a hardware-constrained ER to implement the one-bit feedback in the WR.
As WR is based on a binary tree-structured (TS) codebook, we first elaborate on the binary TS codebook and the WR procedure in Section \ref{SecBT1bitFdBk}.

\subsection{Successive Waveform Refinement with a TS Codebook}\label{SecBT1bitFdBk}
\subsubsection{Binary Tree-Structured Codebook}
\begin{figure}[!t]
\centering
\includegraphics[width = 3.4in]{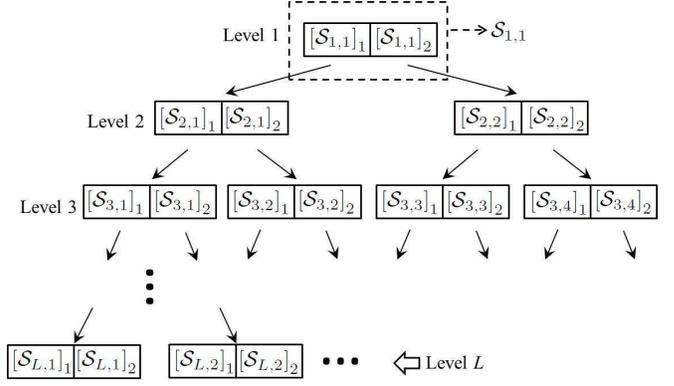}
\caption{Tree-structured codebook $\mathcal{S}$ for waveform refinement.}
\label{FigColocHierCdbk}
\end{figure}
As shown in Fig. \ref{FigColocHierCdbk}, the TS codebook $\mathcal{S}$ consists of $L$ hierarchical levels. The $n_s$\,th subcodebook at level $l$ (for $l = 1,\ldots, L$) is collected into a set $\mathcal{S}_{l,n_s} \subseteq \mathcal{S}$, e.g. $\mathcal{S}_{1,1}$. Subcodebooks at level $l$ have lower resolution than those at level $(l+1)$.  Each level $l$ contains $2^{l-1}$ subcodebooks and $2^l$ codewords. At level $l$, the index $n_s \in \{1,\ldots, 2^{l-1}\}$. Each subcodebook $\mathcal{S}_{l,n_s}$ consists of two codewords $\big[\mathcal{S}_{l,n_s}\big]_1$ and $\big[\mathcal{S}_{l,n_s}\big]_2$. Note that $\big[\mathcal{S}_{l,n_s}\big]_1$ and $\big[\mathcal{S}_{l,n_s}\big]_2$ are column vectors.

Given a $L$-level TS codebook, the preferred highest-resolution codeword (at level $L$) is found by performing \emph{binary searches} throughout the hierarchical levels. The search of a preferred codeword at level $(l+1)$ is confined to the descendent subcodebook (at level $(l+1)$) of the preferred codeword found at level $l$. Fig. \ref{FigColocHierCdbk} characterizes the binary search by arrows: for $l\geq 1$, the $n_s$\,th codeword $\big[ \mathcal{S}_{l,\lceil n_s/2 \rceil} \big]_{n_s - 2(\lceil n_s/2 \rceil-1)}$ at level $l$ is linked with its descendent subcodebook $\mathcal{S}_{l+1,n_s}$ at level $(l+1)$ by a downward arrow, e.g. given $l = 2$ and $n_s = 3$, the 3rd codeword $\big[\mathcal{S}_{2,2}\big]_1$ at level 2 is linked with its descendent subcodebook $\mathcal{S}_{3,3}$ at level 3.

\subsubsection{Waveform Refinement Procedure}
\begin{algorithm}
\caption{WR Procedure with One-Bit Feedback}\label{AlgBeamTrainProtocol}
\begin{algorithmic}[1]
\Statex \textbf{ET Input:} A $L$-level TS codebook $\mathcal{S}$.
\Statex \textbf{ET Output:} A selected precoder $\mathbf{s}^\star$ for WPT.
\State \textbf{Initialization} ET sets $l = 0$, $n_s = 1$ and $f_b = 1$;
\Repeat
    \Statex ET transmitting timeslot:
    \State ET updates $l = l + 1$;
       \begin{equation}
            n_s = \left\{\begin{array}{lr}
            2n_s - 1, & \text{for } f_b = 1 \\
            2n_s, & \text{for } f_b = 0
            \end{array}\right.\nonumber
        \end{equation}
    \State ET sequentially sends two multi-sine signals generated by $\big[\mathcal{S}_{l,n_s}\big]_1$ and $\big[\mathcal{S}_{l,n_s}\big]_2$, respectively;
    \State ER detects and measures $v_{\text{out}, 1}$ and $v_{\text{out}, 2}$ w.r.t. $\big[\mathcal{S}_{l,n_s}\big]_1$ and $\big[\mathcal{S}_{l,n_s}\big]_2$, respectively;
    \Statex ER feedback timeslot:
    \State ER returns one-bit feedback $f_b$ generated by (\ref{EqGeneOneBitFeedback}); \label{AlgBeamTrain_CompFed}
\Until{l = L}
\State ET selects $\mathbf{s}^\star =
        \left\{\begin{array}{lr}
        \big[\mathcal{S}_{l,n_s}\big]_1, & \text{for } f_b = 1 \\
        \big[\mathcal{S}_{l,n_s}\big]_2, & \text{for } f_b = 0
        \end{array}\right.$
\end{algorithmic}
\end{algorithm}
As shown in Fig. \ref{FigBeamTrainProtocol_ColocUsers}, the WR phase is divided into $L$ stages, where each stage consists of ET transmission timeslot 1, ET transmission timeslot 2 and an ER feedback timeslot. The WR procedure boils down to a search over the TS codebook $\mathcal{S}$.
In the ET transmission timeslots at stage $l$, the ET takes two precoders $\big[\mathcal{S}_{l,n_s}\big]_1$ and $\big[\mathcal{S}_{l,n_s}\big]_2$ from a subcodebook $\mathcal{S}_{l,n_s}$ at level $l$ of $\mathcal{S}$ to generate two multi-sine signals, respectively. These two signals are then delivered in ET transmission timeslots 1 and 2, respectively.
Once receiving these two multi-sine signals, the ER measures the weighted sum $v_{\text{out}}$ for these two signals. For notational simplicity, these weighted sum $v_{\text{out}}$ w.r.t. $\big[\mathcal{S}_{l,n_s}\big]_1$ and $\big[\mathcal{S}_{l,n_s}\big]_2$ are designated as $v_{\text{out}, 1}$ and $v_{\text{out}, 2}$, respectively. The ER generates a feedback bit $f_b$ by
\begin{equation}
\label{EqGeneOneBitFeedback}
f_b = \left\{\begin{array}{lr}
        1, & \text{for } v_{\text{out}, 1} > v_{\text{out}, 2} \\
        0, & \text{for } v_{\text{out}, 1} \leq v_{\text{out}, 2}
        \end{array}\right.
\end{equation}
The procedure of the ET transmissions in timeslots 1 and 2 and the above comparison (\ref{EqGeneOneBitFeedback}) boils down to finding a desired codeword that maximizes $v_{\text{out}}$ for a given channel realization, by conducting a search over the subcodebook $\mathcal{S}_{l,n_s}$. From the viewpoint of codebook search, $f_b = 1$ means that $\big[\mathcal{S}_{l,n_s}\big]_1$ is the desired codeword at level $l$, while $f_b = 0$ means that $\big[\mathcal{S}_{l,n_s}\big]_2$ is the desired codeword.
Once the ET receives the feedback\footnote{This feedback can be conducted by conventional communications at the ER. As the feedback bit only indicates an increase or decrease in the harvested energy (or $v_{\text{out}}$) at the ER, it is promising for the ER to implement the feedback of the bit by backscatter communications \cite{BR14}, so as to further reduce any power-consuming processing at the ER.}, it determines which descendent subcodebook should be used to generate multi-sine signals at the $(l+1)$\,th WR stage. According to the binary search rule, if $f_b = 1$, the descendent subcodebook (at level $(l+1)$) of $\big[\mathcal{S}_{l,n_s}\big]_1$ is selected; otherwise, the descendent subcodebook of $\big[\mathcal{S}_{l,n_s}\big]_2$ is selected. The WR procedure is summarized in Algorithm \ref{AlgBeamTrainProtocol}. Conducting binary searches over a TS codebook where the highest level has $N_p$ codewords, the ET delivers $2\log_2 N_p$ multi-sine signals, while the ER returns $\log_2 N_p$ feedback bits.

\subsection{Tree-Structured Codebook Design}\label{SecColocTScdbkDesign}
\subsubsection{TS Codebook Derivation Strategy}
The hierarchical structure and the binary search make the codebook design different from that in Section \ref{SecCdbkDesignAlgoVQ}. The binary search indicates that if $[\mathcal{S}_{l,n_s}]_{n_p}$ is the desired codeword at level $l$, the search of the desired codeword at level $(l+1)$ is confined to the descendent subcodebook of $[\mathcal{S}_{l,n_s}]_{n_p}$.
Meanwhile, we notice that given a finite channel realization set $\mathcal{H}_{0,l,n_s}$ and a subcodebook $\mathcal{S}_{l,n_s}$, searching desired codewords for all elements in $\mathcal{H}_{0,l,n_s}$ is equivalent to dividing $\mathcal{H}_{0,l,n_s}$ into two subsets $\{\mathcal{H}_{0,l,n_s,n_p}\}_{n_p=1}^{2}$. Namely, for a given $n_p$,
\begin{IEEEeqnarray}{rl}
\label{EqTSVQchannelsubset}
\mathcal{H}_{0,l,n_s, n_p} \! = & \Big\{ \! \mathbf{H} \! : f_d \! \Big(\!\mathbf{H}, \mathbf{s}_\text{opt}, [\mathcal{S}_{l,n_s}]_{n_p}\!\Big) \! \leq \! f_d \! \Big(\!\mathbf{H}, \mathbf{s}_\text{opt}, [\mathcal{S}_{l,n_s}]_{n_p^\prime}\!\Big), \nonumber\\
& \forall\, \mathbf{H} \! \in \! \mathcal{H}_{0,l,n_s}, \forall\, \mathbf{s}_\text{opt} \! \in \! \mathcal{S}_{\text{opt},l,n_s}, \forall\, n_p^\prime \neq n_p \Big\}. \quad
\end{IEEEeqnarray}
%
We then illustrate the codebook derivation strategy by the following example.
Given a finite training set $\mathcal{H}_0$ (which is introduced in Section \ref{SecImperfectCSIT}) and the level $1$ subcodebook $\mathcal{S}_{1,1}$, in order to derive the subcodebooks $\mathcal{S}_{2,1}$ and $\mathcal{S}_{2,2}$, we first perform (\ref{EqTSVQchannelsubset}) for $\mathcal{H}_0$ and $\{[\mathcal{S}_{1,1}]_{n_p}\}_{n_p=1}^{2}$ by conducting an exhaustive search over $\mathcal{H}_0$, yielding $\{\mathcal{H}_{0,1,1,n_p}\}_{n_p=1}^{2}$. Then, $\mathcal{S}_{2,1}$ and $\mathcal{S}_{2,2}$, which are the descendants of $[\mathcal{S}_{1,1}]_1$ and $[\mathcal{S}_{1,1}]_2$, can be derived from $\mathcal{H}_{0,1,1,1}$ and $\mathcal{H}_{0,1,1,2}$, respectively. Afterwards, $\mathcal{S}_{3,1}$ (i.e. the descendant of $[\mathcal{S}_{2,1}]_1$) can be derived from the channel realization set $\mathcal{H}_{0,2,1,1}$, which is obtained by (\ref{EqTSVQchannelsubset}) for given $\mathcal{H}_{0,2,1}$ (which is equal to $\mathcal{H}_{0,1,1,1}$) and $\mathcal{S}_{2,1}$. Hence, to sum up, the descendent subcodebook of the codeword $[\mathcal{S}_{l,n_s}]_{n_p}$ (for $l \geq 2$) is derived from its corresponding training set $\mathcal{H}_{0,l,n_s, n_p}$, which is obtained by performing (\ref{EqTSVQchannelsubset}) for given $\mathcal{H}_{0,l,n_s}$ and $\mathcal{S}_{l,n_s}$.

\subsubsection{Designing Level $l$ (for $l \geq 2$) Subcodebook $\mathcal{S}_{l,n_s}$}
Given a codeword $\big[ \! \mathcal{S}_{l\!-\!1,\lceil n_s/2 \rceil} \! \big]_{n_s \! - \! 2(\lceil n_s/2 \rceil \!-\! 1)}$ at level $(l-1)$ and its corresponding training set $\mathcal{H}_{0,l,n_s}$, this part derives the descendent subcodebook $\mathcal{S}_{l,n_s}$ at level $l$ from $\mathcal{H}_{0,l,n_s}$.
In order to evaluate the distortion, we define a set $\mathcal{S}_{\text{opt},l,n_s}$, where each element $\mathbf{s}_\text{opt} \in \mathcal{S}_{\text{opt},l,n_s}$ is computed by \cite[Algorithm 2]{HC16TSParxiv} for a given channel realization $\mathbf{H}^{[t_0]} \in \mathcal{H}_{0,l,n_s}$.
Then, the design problem can be formulated as
\begin{equation}
\label{Prob_TSVQsubCdbk}
\mathcal{S}_{l,n_s}^\star \!=\! \arg \min_{\mathcal{S}_{l,n_s}} \!\left\{\! \bar{f}_d: \left\|[\mathcal{S}_{l,n_s}]_{n_p}\right\|^2 \!\leq\!  P, \, n_p \in \{1,2\} \!\right\}\,,
\end{equation}
where the sum distortion $\bar{f}_d$ is
\begin{IEEEeqnarray}{l}
\label{EqTSVQsubCdbkSumDistort}
\bar{f}_d \!\left( \{\mathcal{H}_{0,l,n_s,n_p}\}_{n_p=1}^{2}, \{\mathcal{S}_{\text{opt},l,n_s,n_p}\}_{n_p=1}^{2}, \mathcal{S}_{l,n_s} \right) = \nonumber \\
\quad  \sum_{n_p = 1}^{2}\, \sum_{\mathbf{H} \in \mathcal{H}_{0,l,n_s,n_p}, \mathbf{s}_\text{opt} \in \mathcal{S}_{\text{opt},l,n_s,n_p} }\! f_d\! \Big( \! \mathbf{H}, \mathbf{s}_\text{opt}, [\mathcal{S}_{l,n_s}]_{n_p} \! \Big). \quad
\end{IEEEeqnarray}
In the above equation, $\{\mathcal{H}_{0,l,n_s,n_p}\}_{n_p=1}^{2}$ is obtained by (\ref{EqTSVQchannelsubset}) for given $\mathcal{S}_{l,n_s}$ and $\mathcal{H}_{0,l,n_s}$.
Additionally, as each element in $\mathcal{S}_{\text{opt},l,n_s}$ is computed for a given element in $\mathcal{H}_{0,l,n_s}$, a set $\mathcal{S}_{\text{opt},l,n_s,n_p} \subseteq \mathcal{S}_{l,n_s}$ corresponding to $\mathcal{H}_{0,l,n_s, n_p}$ can also be found.
The coupled optimization variables make (\ref{Prob_TSVQsubCdbk}) intractable. To solve (\ref{Prob_TSVQsubCdbk}), we regard $\{\mathcal{H}_{0,l,n_s,n_p}\}_{n_p=1}^{2}$ and $\mathcal{S}_{l,n_s}$ as independent optimization variables, similarly to Section \ref{SecCdbkDesignAlgoVQ_AlgDesign}.
It can be seen that problem (\ref{Prob_TSVQsubCdbk}) has the same form as problem (\ref{Prob_VQ}). Intuitively, (\ref{Prob_TSVQsubCdbk}) can be solved by Algorithm \ref{AlgVQcdbk}. Therefore,
\begin{equation}
\label{AlgTSVQsubCdbk}
\mathcal{S}_{l,n_s}^\star = {\bf\Phi}\big(\mathcal{H}_{0,l,n_s}, \mathcal{S}_{\text{opt},l,n_s}, \mathcal{S}^{(0)}_{l,n_s}, 2 \big)\,.
\end{equation}
However, given an \emph{arbitrary} initial subcodebook $\mathcal{S}^{(0)}_{l,n_s}$, the optimized $\mathcal{S}_{l,n_s}^\star$ offered by (\ref{AlgTSVQsubCdbk}) may not satisfy the necessary condition for an optimized subcodebook $\mathcal{S}_{l,n_s}^\star$:
\begin{IEEEeqnarray}{l}
\label{EqNeceCondSubCdbkTScdbk}
\bar{f}_d \!\left( \{\mathcal{H}_{0,l,n_s,n_p}^\star\}_{n_p=1}^{2}, \{\mathcal{S}_{\text{opt},l,n_s,n_p}^\star\}_{n_p=1}^{2}, \mathcal{S}_{l,n_s}^\star \right) \leq \nonumber\\
\quad \bar{f}_d \!\left( \mathcal{H}_{0,l,n_s}, \mathcal{S}_{\text{opt},l,n_s}, \big[ \mathcal{S}_{l-1,\lceil n_s/2 \rceil} \big]_{n_s - 2(\lceil n_s/2 \rceil-1)} \right), \quad
\end{IEEEeqnarray}
where $ \{\mathcal{H}_{0,l,n_s,n_p}^\star\}_{n_p=1}^{2}$ is obtained by (\ref{EqTSVQchannelsubset}) for given $\mathcal{S}_{l,n_s}^\star$ and $\mathcal{H}_{0,l,n_s}$.
Recall that when performing Algorithm \ref{AlgVQcdbk}, $\bar{f}_d$ monotonically decreases over iterations.
Therefore, in order to make the optimized $\mathcal{S}_{l,n_s}^\star$ always satisfy (\ref{EqNeceCondSubCdbkTScdbk}), the initial $\mathcal{S}^{(0)}_{l,n_s}$ has to satisfy
\begin{IEEEeqnarray}{l}
\label{EqNeceCondIniSubCdbkTScdbk}
\bar{f}_d \!\left( \{\mathcal{H}_{0,l,n_s,n_p}^\prime\}_{n_p=1}^{2}, \{\mathcal{S}_{\text{opt},l,n_s,n_p}^\prime\}_{n_p=1}^{2}, \mathcal{S}^{(0)}_{l,n_s} \right) \leq \nonumber\\
\quad \bar{f}_d \!\left( \mathcal{H}_{0,l,n_s}, \mathcal{S}_{\text{opt},l,n_s}, \big[ \mathcal{S}_{l-1,\lceil n_s/2 \rceil} \big]_{n_s - 2(\lceil n_s/2 \rceil-1)} \right), \quad
\end{IEEEeqnarray}
where $\{\mathcal{H}_{0,l,n_s,n_p}^\prime\}_{n_p=1}^{2}$ is obtained by (\ref{EqTSVQchannelsubset}) for given $\mathcal{S}^{(0)}_{l,n_s}$ and $\mathcal{H}_{0,l,n_s}$.
\begin{algorithm}
\caption{Subcodebook Initialization}\label{AlgIntializeSubCdbk_TS}
\begin{algorithmic}[1]
\Statex \textbf{Input:} $\mathcal{H}_{0,l,n_s}$, $\mathcal{S}_{\text{opt},l,n_s}$ and $\big[ \mathcal{S}_{l-1,\lceil n_s/2 \rceil} \big]_{n_s - 2(\lceil n_s/2 \rceil-1)}$, where $l \geq 2$; the $t_0$\, th elements within $\mathcal{H}_{0,l,n_s}$ and $\mathcal{S}_{\text{opt},l,n_s}$ are designated as $\mathbf{H}^{[t_0]}$ and $\mathbf{s}_{\text{opt}}^{[t_0]}$, respectively.
\Statex \textbf{Output:} $\mathcal{S}^{(0)}_{l,n_s}$;
\State \textbf{Initialization} $t_0  = 0$;
\State $[\mathcal{S}_{l,n_s}]_1^{(0)} = \big[ \mathcal{S}_{l-1,\lceil n_s/2 \rceil} \big]_{n_s - 2(\lceil n_s/2 \rceil-1)}$;
\Repeat
    \State $t_0  = t_0 + 1$;
    \State $f_{d,\min} = f_d \big(\mathbf{H}^{[t_0]}, \mathbf{s}_{\text{opt}}^{[t_0]}, [\mathcal{S}_{l,n_s}]_1^{(0)} \big)$;
\Until{f_{d,\min} \geq \epsilon}
\State $[\mathcal{S}_{l,n_s}]_2^{(0)} = \mathbf{s}_{\text{opt}}^{[t_0]}$;
\end{algorithmic}
\end{algorithm}
\begin{algorithm}
\caption{TS Codebook Design Algorithm}\label{AlgTSVQcdbk}
\begin{algorithmic}[1]
\Statex \textbf{Input:} $\mathcal{H}_0$, $\mathcal{S}_\text{opt}$, $\mathbf{s}_\text{int}$ and $L$.
\Statex \textbf{Output:} The TS codebook $\mathcal{S}$.
\State \textbf{Initialization} $\mathcal{S}_0 = {\bf\Psi} \left(\mathcal{H}_0, \mathbf{s}_\text{int}\right)$;
\State \textbf{for} $l = 1:1:L$
\State \quad \textbf{for} $n_s = 1:1:2^{l-1}$
\State \quad\quad \textbf{if} $l = 1$
\State \quad\quad\quad $\mathcal{S}^{(0)}_{1,1} = {\bf\Theta}\big(\mathcal{H}_0 , \mathcal{S}_\text{opt}, \mathcal{S}_0\big)$;
\State \quad\quad\quad $\mathcal{S}_{1,1}^\star = {\bf\Phi}\big(\mathcal{H}_0, \mathcal{S}_\text{opt}, \mathcal{S}^{(0)}_{1,1}, 2 \big)$;
\State \quad\quad\quad Update $\{\mathcal{H}_{0,1,1,n_p}\}_{n_p=1}^{2}$ by (\ref{EqTSVQchannelsubset}) with $\mathcal{S}_{1,1}$; update $\{\mathcal{S}_{\text{opt},1,1,n_p}\}_{n_p=1}^{2}$ corresponding to $\{\mathcal{H}_{0,1,1,n_p}\}_{n_p=1}^{2}$;
\State \quad\quad \textbf{else}
\State \quad\quad\quad $l^\prime = l-1$; $n_s^\prime = \lceil \frac{n_s}{2} \rceil$; $n_p^\prime = n_s \!-\! 2(\lceil \frac{n_s}{2} \rceil \!-\! 1)$;
\State \quad\quad\quad $\mathcal{H}_{0,l,n_s} = \mathcal{H}_{0, l^\prime, n_s^\prime, n_p^\prime}$ and $\mathcal{S}_{\text{opt},l,n_s} = \mathcal{S}_{\text{opt},l^\prime,n_s^\prime,n_p^\prime}$;
\State \quad\quad\quad $\mathcal{S}^{(0)}_{l,n_s}  = {\bf\Theta}\bigg(\mathcal{H}_{0,l,n_s}, \mathcal{S}_{\text{opt},l,n_s}, \Big[ \mathcal{S}_{l^\prime, n_s^\prime} \Big]_{n_p^\prime} \bigg)$;
\State \quad\quad\quad $\mathcal{S}_{l,n_s}^\star = {\bf\Phi}\big(\mathcal{H}_{0,l,n_s}, \mathcal{S}_{\text{opt},l,n_s}, \mathcal{S}^{(0)}_{l,n_s}, 2 \big)$;
\State \quad\quad\quad Update $\{\mathcal{H}_{0,l,n_s,n_p}\}_{n_p=1}^{2}$ by (\ref{EqTSVQchannelsubset}) with $\mathcal{S}_{1,1}$; update $\{\mathcal{S}_{\text{opt},l,n_s,n_p}\}_{n_p=1}^{2}$ corresponding to $\{\mathcal{H}_{0,l,n_s,n_p}\}_{n_p=1}^{2}$;
\State \quad\quad \textbf{end}
\State \quad \textbf{end}
\State \textbf{end}
\end{algorithmic}
\end{algorithm}
As the elements within each $\mathcal{H}_{0,l,n_s,n_p}$ are taken from $\mathcal{H}_{0,l,n_s}$, it can be shown by (\ref{EqTSVQchannelsubset}) that if one codeword in $\mathcal{S}^{(0)}_{l,n_s}$ is set as $\big[ \mathcal{S}_{l-1,\lceil n_s/2 \rceil} \big]_{n_s - 2(\lceil n_s/2 \rceil-1)}$, (\ref{EqNeceCondIniSubCdbkTScdbk}) can always be satisfied.
Specifically, assuming that the other codeword within $\mathcal{S}^{(0)}_{l,n_s}$ is $\big[\mathcal{S}^{(0)}_{l,n_s}\big]_2$, if all the channel realizations within $\mathcal{H}_{0,l,n_s}$ can achieve a smaller $f_d$ with $\big[ \mathcal{S}_{l-1,\lceil n_s/2 \rceil} \big]_{n_s - 2(\lceil n_s/2 \rceil-1)}$ than $\big[\mathcal{S}^{(0)}_{l,n_s}\big]_2$, equality holds in (\ref{EqNeceCondIniSubCdbkTScdbk}); otherwise, the left hand side of (\ref{EqNeceCondIniSubCdbkTScdbk}) is less than the right hand side of (\ref{EqNeceCondIniSubCdbkTScdbk}).
Therefore, for level $l \geq 2$, we design an $\mathcal{S}^{(0)}_{l,n_s}$ initialization algorithm, as shown in Algorithm \ref{AlgIntializeSubCdbk_TS}. In the following discussion, we express the initialization algorithm as a function
\begin{equation}
\label{EqFuncAlgIntializeSubCdbk_TS}
\mathcal{S}^{(0)}_{l,n_s}\! =\! {\bf\Theta}\bigg(\mathcal{H}_{0,l,n_s}\!, \mathcal{S}_{\text{opt},l,n_s}\!, \Big[ \mathcal{S}_{l-1,\lceil \frac{n_s}{2} \rceil} \Big]_{n_s \!-\! 2(\lceil \frac{n_s}{2} \rceil \!-\! 1)} \bigg).
\end{equation}

\subsubsection{Designing Level $1$ Subcodebook $\mathcal{S}_{1,1}$}
To initialize the subcodebook at level 1, we first derive a root codeword $\mathcal{S}_0$ from the entire training set $\mathcal{H}_0$ by Algorithm \ref{AlgSAA}, i.e.  $\mathcal{S}_0 = {\bf\Psi} \left(\mathcal{H}_0, \mathbf{s}_\text{int}\right)$. This root codeword $\mathcal{S}_0$ maximizes the expected weighted sum-$v_\text{out}$ for $\mathcal{H}_0$. The initial subcodebook $\mathcal{S}^{(0)}_{1,1}$ at level 1 can be generated by substituting the root codeword $\mathcal{S}_0$ into Algorithm \ref{AlgIntializeSubCdbk_TS}, i.e. $\mathcal{S}^{(0)}_{1,1} = {\bf\Theta}\big(\mathcal{H}_0 , \mathcal{S}_\text{opt}, \mathcal{S}_0\big)$. Then, $\mathcal{S}_{1,1}$ can be achieved by performing $\mathcal{S}_{1,1}^\star = {\bf\Phi}\big(\mathcal{H}_0, \mathcal{S}_\text{opt}, \mathcal{S}^{(0)}_{1,1}, 2 \big)$.

The proposed TS codebook design algorithm is summarized in Algorithm \ref{AlgTSVQcdbk}. The necessary condition (\ref{EqNeceCondSubCdbkTScdbk}) implies that the subcodebook $\mathcal{S}_{l,n_s}^\star$ optimized by solving (\ref{Prob_TSVQsubCdbk}) essentially minimizes the average distortion only for the channel realization set $\mathcal{H}_{0,l,n_s}$, but not the entire training set $\mathcal{H}_0$.
From this, it can be inferred that the highest-resolution codewords at the highest level of the TS codebook are not optimized to minimize the average distortion for the entire training set $\mathcal{H}_0$. Therefore, compared to the codebook for WS-based WPT, the TS codebook is suboptimal for maximizing the average $v_\text{out}$.

\section{Performance Evaluations}\label{SecSimResults}
We consider spatially-uncorrelated frequency-selective fading channels at a central frequency of $2.4$\,GHz, where the channels are generated by the IEEE TGn NLOS channel model E\cite{ESetal04}. This model characterizes a typical large open space indoor (or outdoor) WiFi-like environment. The path loss, which is used to compute the large-scale fading $\Lambda^{1/2}$, is modeled according to \cite{ESetal04}, under the assumption of 0dB transmit and receive antenna gains. In the codebook designs, we set $w_q = 1 \, \forall q \in\{1, \ldots, Q\}$. The Equivalent Isotropically Radiated Power (EIRP, which is defined as the product of $MP$ at the ET), the channel bandwidth and the path loss are respectively set as $36$\,dBm, $10$\,MHz and $60.046$\,dB (which corresponds to a distance $D=10$\,m), unless otherwise stated.
Please note that the nonlinear rectenna model holds primarily in the low input power regime ($-30$\,dBm to $0$\,dBm).
In the simulations, we assume that the ER feedback is errorless.

\subsection{Waveform Selection-Based WPT}
\begin{figure}[!t]
\centering
\includegraphics[width = 3.4in]{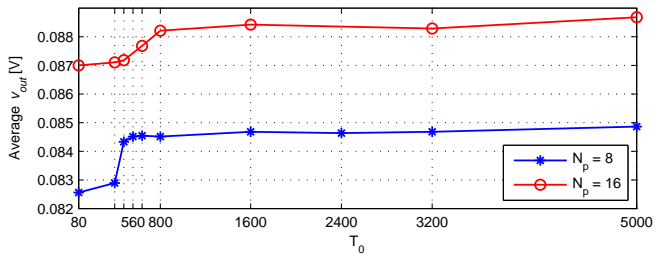}
\caption{Average $v_\text{out}$ as a function of $T_0$, with $M=1$, $N=8$ and $Q=1$.}
\label{Fig_AvgVout_M1N8_vs_T0}
\end{figure}
We first draw the effect of the number of channel realizations in a training set and the number of codewords on the average $v_\text{out}$. Fig. \ref{Fig_AvgVout_M1N8_vs_T0} investigates the average $v_\text{out}$ in the WPT phase as a function of $T_0$, in the presence of various $N_p$.
It is shown that the average $v_\text{out}$ achieved by an $N_p$-codeword codebook increases as $T_0$ increases, but saturates at a high $T_0$.
Moreover, in the presence of a larger $N_p$, the average $v_\text{out}$ does not saturate until a higher $T_0$.
It is remarkable that although the per-iteration (asymptotic) complexity of the SAA algorithm (which is exploited to solve the codewords optimization subproblem) is not a function of $T_0$, increasing $T_0$ causes an increase in the elapsed time of the partition optimization. Therefore, the elapsed time of Algorithm \ref{AlgVQcdbk} increases.

\begin{figure}[!t]
\centering
\includegraphics[width = 3.4in]{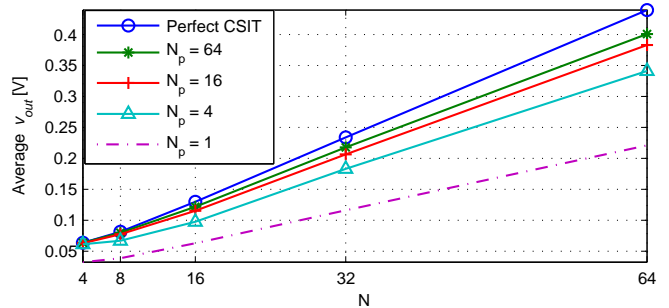}
\caption{Average $v_\text{out}$ in the WPT phase as a function of $N$, with $M = 1$ and $Q = 1$.}
\label{Fig_Vout_vs_N_diff_Np_LossScaleLaw}
\end{figure}

Fig. \ref{Fig_Vout_vs_N_diff_Np_LossScaleLaw} investigates the average $v_\text{out}$ in the WPT phase as a function of $N$, for various $N_p$ (which is a function of the number of feedback bits).
In the simulations, codebooks at the ET are derived with a training set $\mathcal{H}_0$ of $T_0 = 5000$ channel realizations. In the presence of perfect CSIT, $v_\text{out}$ is maximized by performing \cite[Algorithm 1]{HC16TSParxiv} to optimize ET transmit waveforms.
It can be seen that as $N$ increases\footnote{It is shown that given $P = 36$dBm, the average $v_\text{out}$ always scales with the increasing $N$ but does not saturate or decrease. This is due to that the derivation (in \cite{HC16TSParxiv, HC16SPAWC}) of the nonlinear rectenna model (\ref{EqVoutq_Scalar2}) assumes an ideal low pass filter. Hence, as $N$ increases, the input waveform peak increases. Nevertheless, if $N$ kept growing, the waveform peak would finally have a very high amplitude, and the diode would be biased in the resistive zone and present a linear I-V characteristic. In such a scenario, the nonlinear model is inapplicable\cite{BCCG13}. Please note that according to the output DC voltage (as a function of input power) observed in measurements and circuit simulations\cite{BBFCGC15, BC11}, the nonlinear model is valid throughout the MATLAB simulations conducted in the paper. Additionally, in order to exploit the output DC power gain offered by multi-sine WPT in practice, as a part of a systematic treatment, the matching network, the low pass filter and the load impedance should be jointly optimized for the rectenna design\cite{CB16, CB17arXiv, KCM17}.},
the gap between the average $v_\text{out}$ achieved with perfect CSIT and the average $v_\text{out}$ achieved by the WS-based WPT for a given $N_p$ grows as $N$ increases. This observation indicates that to guarantee a certain voltage loss, the number of feedback bits should scale with $N$.
It is also demonstrated that similarly to WPT with perfect CSIT, the WS-based WPT can enhance the average $v_\text{out}$ by exploiting more sinewaves, where the average $v_\text{out}$ scales linearly with the increasing $N$.
Additionally, Fig. \ref{Fig_Vout_vs_N_diff_Np_LossScaleLaw} indicates that increasing the number of feedback bits enables a faster increase in the average $v_\text{out}$ as a function of the number of sinewaves. This can be drawn from the observation that a growing $N_p$ can increase the slope of the average $v_\text{out}$ curve w.r.t. the WS-based WPT.

\begin{figure}[!t]
\centering
\includegraphics[width = 3.4in]{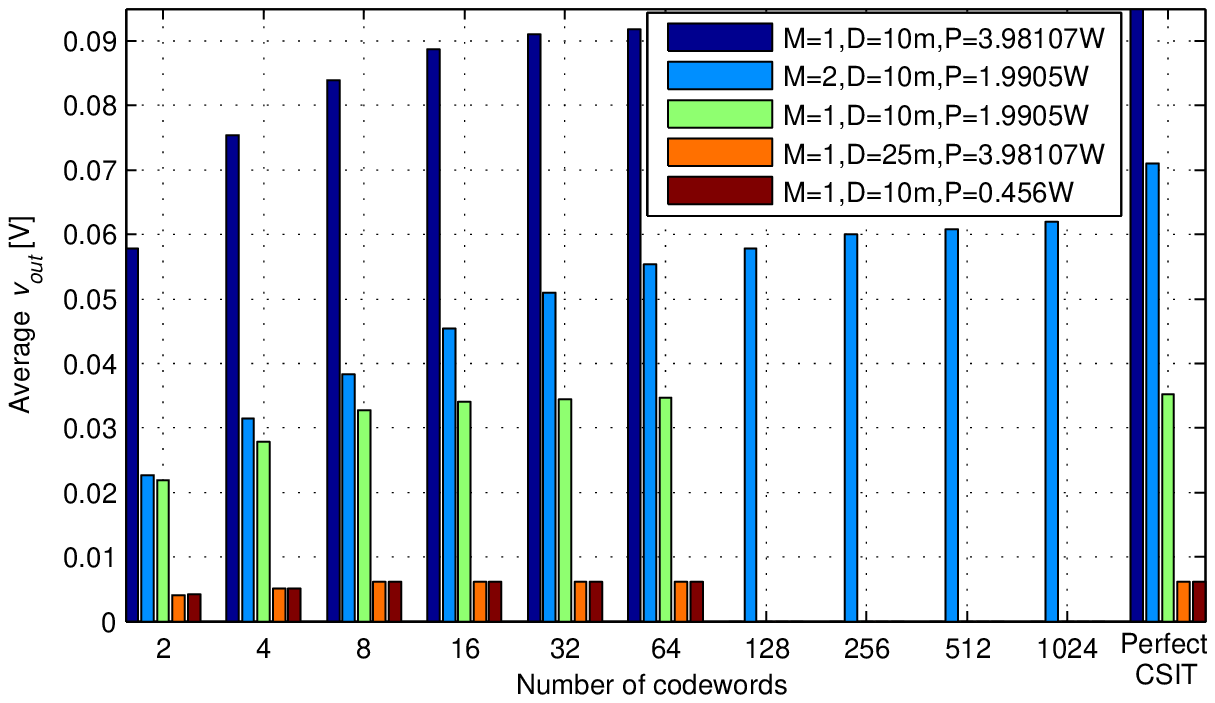}
\caption{Average $v_\text{out}$ in the WPT phase as a function of $N_p$, with $N = 8$ and $Q = 1$.}
\label{Fig_Vout_vs_Np_Effect_of_M_Lambda_P}
\end{figure}
Fig. \ref{Fig_Vout_vs_Np_Effect_of_M_Lambda_P} studies the effect of $M$, $\Lambda$ and $P$ on the average $v_\text{out}$ achieved by the WS-based WPT, where codebooks for various $N_p$, $M$, $P$ and $D$ are constructed. In the legend of Fig. \ref{Fig_Vout_vs_Np_Effect_of_M_Lambda_P}, $D = 25$\,m corresponds to path loss of $69.4584$\,dB.
We draw the following observations from Fig. \ref{Fig_Vout_vs_Np_Effect_of_M_Lambda_P}.
1) The average $v_\text{out}$ offered by the setups of ``$M=1$, $D=25$\,m, $P=3.98107$\,W" and ``$M=1$, $D=10$\,m, $P=0.456$\,W" illustrates that in the presence of a small average input power into the rectenna, the average $v_\text{out}$ as a function of $N_p$ quickly saturates, even if $N_p$ is small. Due to the similar $\Lambda P$ values\footnote{Note that $\Lambda$ is related to the large-scale fading $\Lambda^{1/2}$, which is a function of the WPT transmission distance $D$.}, these two setups yield similar average $v_\text{out}$.
2) Fig. \ref{Fig_Vout_vs_Np_Effect_of_M_Lambda_P} reveals that the average $v_\text{out}$ achieved by the WS-based WPT for a given number of feedback bits (or $N_p$) can be improved by increasing $M$, $P$ or $\Lambda$. Such an operation improves the average input power into the rectenna.
Moreover, the $v_\text{out}$ achieved by the setups of ``$M=1$, $D=10$\,m, $P=1.9905$\,W" and ``$M=2$, $D=10$\,m, $P=1.9905$\,W" indicates that a small $N_p$ can become a limiting factor, such that increasing $M$ may not lead to a significant increase in the average $v_\text{out}$.
Note that this observation is quite different from the observation made in the presence of perfect CSIT, where increasing $M$ significantly enhances the average $v_\text{out}$\cite{CB16, HC16TSParxiv}.
3) In the scenario where $M$, $P$ or $\Lambda$ is enlarged, the number of feedback bits (or $N_p$) should be increased, so as to guarantee a certain voltage loss. Simulation results imply that for $M \geq 2$, a large $N_p$ is necessary for the WS-based WPT to reduce the average $v_\text{out}$ performance loss.

\begin{figure}[!t]
\centering
\includegraphics[width = 3.4in]{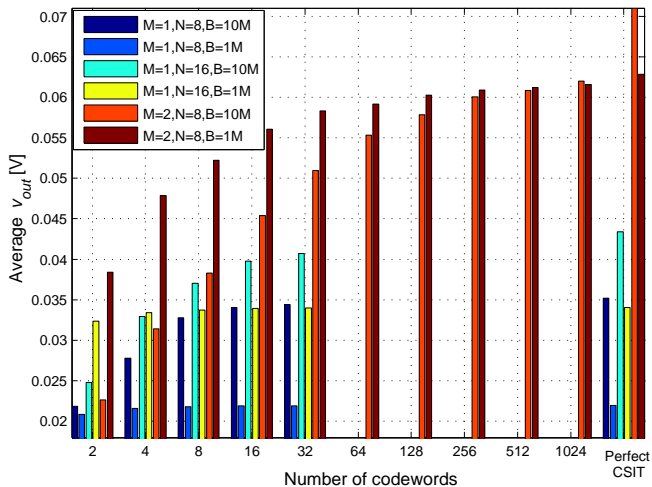}
\caption{Average $v_\text{out}$ in the WPT phase as a function of $N_p$, with $N=8$, $Q = 1$ and $P=1.9905$\,W.}
\label{Fig_Vout_vs_Np_B10M_1M}
\end{figure}
\begin{figure}[!t]
\centering
\includegraphics[width = 2.4in]{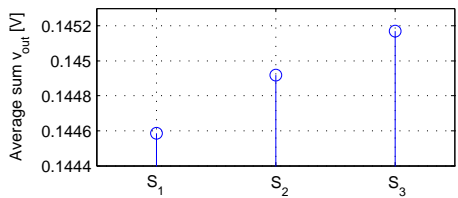}
\caption{Average sum $v_\text{out}$ achieved with codebooks $\mathcal{S}_1$, $\mathcal{S}_2$ and $\mathcal{S}_3$, with $M = 1$, $N = 8$ and $N_p = 8$.}
\label{Fig_AvgSumVout_Cdbk_S1NoU1D5k_S2NoU1D10k_S3NoU2D5k}
\end{figure}
Fig. \ref{Fig_Vout_vs_Np_B10M_1M} studies the effect of transmission bandwidth $B$ on the average $v_\text{out}$ achieved by the WS-based WPT, where codebooks are designed for various $N_p$, $M$, $N$ and $B$. Note that in the simulations, a channel with $B=10$\,MHz is much more frequency-selective than that with $B=1$\,MHz.
Fig. \ref{Fig_Vout_vs_Np_B10M_1M} shows that for given $M$ and $N$, a large bandwidth always leads to a higher average $v_\text{out}$ than a small bandwidth in the presence of perfect CSIT, thanks to the frequency selectivity gain.
By contrast, it is shown that performing limited feedback-based WPT with multiple transmit antennas (e.g. $M = 2$ and $N = 8$) or a large number of sinewaves (e.g. $M = 1$ and $N = 16$), a large bandwidth may offer a lower average $v_\text{out}$ than a small bandwidth in the presence of a small $N_p$, although the former outperforms the latter voltage-wise in the presence of a large $N_p$. This stems from the fact that a small bandwidth features little frequency selectivity and enables a fast increase in the average $v_\text{out}$ as a function of $N_p$.
Fig. \ref{Fig_Vout_vs_Np_B10M_1M} suggests that in the presence of multiple transmit antennas or a large number of sinewaves, WPT based on limited feedback can exploit a small transmission bandwidth to alleviate a high overhead caused by a long WS phase, if the output voltage performance loss is affordable.

We now draw insights into the codebook design for a multi-rectenna ER.
In the codebook design for a \emph{$Q$-rectenna} ER, the accuracy of the codeword $[\mathcal{S}]_{n_p}$ obtained by solving the codewords optimization (\ref{Prob_VQ_CentroidCondition_equi}) for a given partition cell $\mathcal{H}_{0,n_p}^\prime$ is essentially determined by the total number of channel gain vectors $\mathbf{h}_q^{[t_0]}$ within $\mathcal{H}_{0,n_p}^\prime$, which is equal to $T_{n_p}^\prime Q$ (where $T_{n_p}^\prime \triangleq \text{Card}(\mathcal{H}_{0,n_p}^\prime)$). Intuitively, in the codebook design for a \emph{single-rectenna} ER, assuming a partition cell $\mathcal{H}_{0,n_p}^{\prime\prime}$ consisting of $T_{n_p}^\prime Q$ elements i.e. $\text{Card}(\mathcal{H}_{0,n_p}^{\prime\prime}) = T_{n_p}^\prime Q$, solving (\ref{Prob_VQ_CentroidCondition_equi}) for $\mathcal{H}_{0,n_p}^{\prime\prime}$ may yield a codeword similar to that achieved by solving (\ref{Prob_VQ_CentroidCondition_equi}) for $\mathcal{H}_{0,n_p}^\prime$.
A question then arises: how is the $v_\text{out}$ performance of serving a \emph{$Q$-rectenna} ER with a codebook derived for a \emph{single-rectenna} ER?
To find the answer, the simulations for Fig. \ref{Fig_AvgSumVout_Cdbk_S1NoU1D5k_S2NoU1D10k_S3NoU2D5k} consider a scenario where an ET serves a 2-rectenna ER, while codebooks $\mathcal{S}_1$, $\mathcal{S}_2$ and $\mathcal{S}_3$ are derived with parameters $(Q=1, T_0=5000)$, $(Q=1, T_0=10000)$ and $(Q=2, T_0=5000)$, respectively.
Fig. \ref{Fig_AvgSumVout_Cdbk_S1NoU1D5k_S2NoU1D10k_S3NoU2D5k} depicts that codebooks derived for $Q=1$ are suboptimal for serving a $Q$-rectenna ER, achieving lower average $v_\text{out}$ than a codebook derived for a $Q$-rectenna ER. The comparison of the $v_\text{out}$ achieved by $\mathcal{S}_2$ and that achieved by $\mathcal{S}_3$ reveals that the partition optimization (\ref{Eq_VQ_H0np_Update}) shows significant effect on the codebook design for a multi-rectenna ER.

\subsection{Waveform Refinement-Based WPT}
\begin{figure}[!t]
\centering
\includegraphics[width = 3.4in]{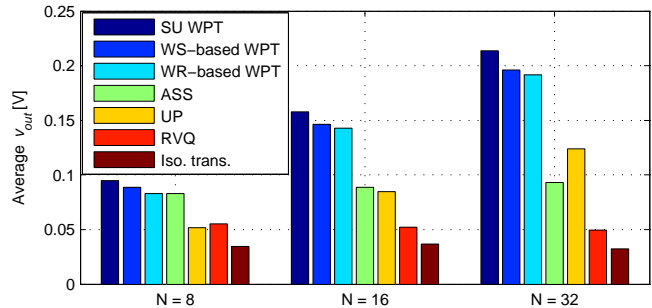}
\caption{Average $v_\text{out}$ in the WPT phase as a function of $N$, with $M=1$ and $Q=1$. For the WS-based WPT and RVQ, the codebook size is set as $N_p = 2N$. In the WR-based WPT, the TS codebook has $L = \log_2 2N$ levels.}
\label{Fig_AvgVout_SUWPT_VQ_TS_ASS_UP_NoCSIT}
\end{figure}
Fig. \ref{Fig_AvgVout_SUWPT_VQ_TS_ASS_UP_NoCSIT} investigates the average $v_\text{out}$ (in the WPT phase) achieved by the proposed waveform strategies relying on limited feedback and baselines.
Simulated with \cite[Algorithm 1]{HC16TSParxiv}, SU WPT presents the average $v_\text{out}$ of multi-sine WPT in the presence of perfect CSIT.
Assuming perfect CSIT, Adaptive Single Sinewave (ASS) \cite{CB16} allocates power to the frequency corresponding to the strongest frequency-domain channel gain. ASS maximizes the harvested power for the WPT waveform optimization under the assumption of the linear rectenna model.
The Uniform Power (UP) allocation scheme assumes perfect CSI on the normalized spatial domain channel but no CSI on the frequency domain channel power gains. Hence, transmit power is equally allocated across frequencies by UP.
Similarly to the WS-based WPT, the RVQ scheme enables the ET to select a codeword from an RVQ codebook for WPT, based on the ER feedback. This RVQ codebook consists of $N_p$ codewords (which are $MN \times 1$ complex vectors) randomly selected from the uniform distribution on the complex unit sphere.
Assuming no CSIT, the isotropic transmission (iso. trans.) uniformly allocates power across frequencies and randomly generates the phases of the transmitted sinewaves.
Fig. \ref{Fig_AvgVout_SUWPT_VQ_TS_ASS_UP_NoCSIT} demonstrates that the average $v_\text{out}$ (in the WPT phase) achieved by the WR-based WPT can be slightly lower than that achieved by the WS-based WPT, although the TS codebook is suboptimal in terms of the average $v_\text{out}$.
It is also shown that WPT strategies based on WR and WS always offer higher average $v_\text{out}$ than UP. This observation confirms that the proposed waveform designs relying on limited feedback are beneficial to multi-sine WPT.
Moreover, the WR-based WPT can significantly outperform ASS in terms of the average $v_\text{out}$.
In other words, the proposed waveform strategies, based on limited feedback, can outperform the linear rectenna model-based waveform design which even assumes perfect CSIT.
Additionally, RVQ is significantly outperformed by the WS/WR-based WPT, in terms of the average $v_\text{out}$ in the WPT phase.

\begin{figure}[!t]
\centering
\includegraphics[width = 3.4in]{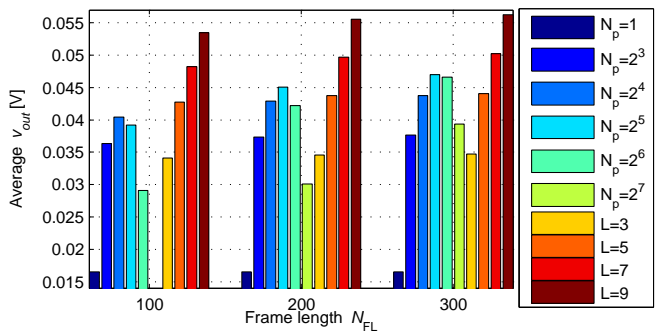}
\caption{Average $v_\text{out}$ (in the entire time frame consisting of the WS/WR phase and the WPT phase) as a function of frame length, with $M=2$, $N=8$ and $Q=1$. The data with parameter $N_p$ is $\bar{v}_\text{out,WS}$, while that with parameter $L$ is $\bar{v}_\text{out,WR}$.}
\label{Fig_AvgVout_vs_FrameLength}
\end{figure}
We now investigate the impact of the duration of the WS (or the WR) phase on the average $v_\text{out}$ in the entire time frame consisting of the WPT phase and the WS (or the WR) phase.
Insights are drawn from the observations made in Fig. \ref{Fig_AvgVout_vs_FrameLength}.
Assumptions made in the simulations are listed as follows. An ET transmission timeslot (as shown in Fig. \ref{FigBeamSelectionProtocol}) in the WS-based WPT has the same duration as an ET transmission timeslot (as shown in Fig. \ref{FigBeamTrainProtocol_ColocUsers}) in the WR-based WPT. This duration is defined as the base unit of time i.e. 1, while the duration of each ER feedback (which is assumed to be errorless) in the WS/WR-based WPT is negligible. The time frames of the WS-based WPT and the WR-based WPT have the same duration $N_\text{FL}$. Wireless channels remain constant in a time frame but varies over different frames. Given a channel realization, the rectenna DC output voltage measured by the ER within the WS or the WR phase is designated as $v_\text{out}(\mathbf{s})$, where $\mathbf{s}$ is the precoder exploited by the ET.
Under the above assumptions and definitions, performing the WS-based WPT for a channel realization, the rectenna DC output voltage $v_\text{out}$ averaged over a time frame can be obtained by
\begin{equation}
v_\text{out,WS} = \frac{\textstyle{\sum_{n_p=1}^{N_p}}\! v_\text{out}\big(\mathcal{S}_{n_p}\big) \!+\! (N_\text{FL} \!-\! N_p)v_\text{out,WS-WPT}}{N_\text{FL}},
\end{equation}
where $v_\text{out,WS-WPT} = \max_{n_p} v_\text{out}\big(\mathcal{S}_{n_p}\big)$ is the $v_\text{out}$ achieved in the WPT phase. Similarly, performing the WR-based WPT for a channel realization, the $v_\text{out}$ averaged over a time frame is
\begin{equation}
v_\text{out,WR} \!=\! \frac{\textstyle{\sum_{l=1}^{L}}\! \textstyle{\sum_{n_p=1}^2} \! v_\text{out}\big(\! \big[\mathcal{S}_{l,n_s^\prime}\big]_{n_p}\! \big) \!+\! (N_\text{FL} \!-\! 2L) v_\text{out,WR-WPT}}{N_\text{FL}},
\end{equation}
where $v_\text{out,WR-WPT} = \max_{n_p\in\{1,2\}} v_\text{out}\big(\big[\mathcal{S}_{L,n_s}\big]_{n_p}\big)$ is the $v_\text{out}$ achieved in the WPT phase, and the precoders $\big[\mathcal{S}_{l,n_s^\prime}\big]_{n_p}$ and $\big[\mathcal{S}_{L,n_s}\big]_{n_p}$ are determined by the WR procedure Algorithm \ref{AlgBeamTrainProtocol}.
In the simulations, $v_\text{out,WS}$ and $v_\text{out,WR}$ are averaged over 300 time frames, and the average values are designated as $\bar{v}_\text{out,WS}$ and $\bar{v}_\text{out,WR}$, respectively. Fig. \ref{Fig_AvgVout_vs_FrameLength} studies $\bar{v}_\text{out,WS}$ and $\bar{v}_\text{out,WR}$ as a function of $N_\text{FL}$. We make the following observations.
1) The special case of the WS-based WPT, where $N_p = 1$ and the WS procedure is eliminated, does not suffer from the overhead of the WS phase. However, such a WPT scheme can be significantly outperformed by other WS-based and WR-based WPT schemes, in terms of the average $v_\text{out}$ in the entire time frame.
2) Not only the overhead caused by the WS/WR phase but also the resolution of the codebook can affect the average $v_\text{out}$ during the entire time frame. This can be drawn from the observation that $\bar{v}_\text{out,WS}$ for $N_p = 2^3$ slightly increases as $N_\text{FL}$ grows, due to a low-resolution codebook. By contrast, $\bar{v}_\text{out,WS}$ for $N_p = 2^6$ soars with the increasing $N_\text{FL}$.
Moreover, it can be drawn that that increasing $N_\text{FL}$ can improve $\bar{v}_\text{out,WS}$ of the WS-based WPT, where the increase in $\bar{v}_\text{out,WS}$ results from a tradeoff between the resolution of the codebook and the length of the WS phase. This can be seen from the observation that when $N_\text{FL}$ is increased from $100$ to $200$, the WS-based WPT benefits from a higher-resolution codebook (whose $N_p = 2^5$), achieving a higher $\bar{v}_\text{out,WS}$. However, for $N_\text{FL} = 200$ and $N_\text{FL} = 300$, the $\bar{v}_\text{out,WS}$ values summit at the same $N_p = 2^5$. This illustrates that the increase in $\bar{v}_\text{out,WS}$ comes from a decrease in the proportion of the overhead.
3) Although the TS codebook for the WR-based WPT is suboptimal for maximizing the average $v_\text{out}$ in the WPT phase, the WR-based WPT can offer a higher average $v_\text{out}$ in the entire time frame, if the WR can significantly reduce the overhead\footnote{Recall that the ET transmits $N_p$ energy signals in the WS phase, while the ET transmits only $2\log_2 N_p$ energy signals in the WR phase (if the highest level of a TS codebook contains $N_p$ codewords).}.
This can be seen from Fig. \ref{Fig_AvgVout_vs_FrameLength} that for $L=3$ and $N_p = 2^3$, $\bar{v}_\text{out,WS}$ is always higher than $\bar{v}_\text{out,WR}$, although the overhead caused by the WR phase is a little lower than that caused by the WS phase. However, for $L=7$ and $N_p = 2^7$, $\bar{v}_\text{out,WR}$ is much higher than $\bar{v}_\text{out,WS}$\footnote{For $L=7$ and $N_p = 2^7$, in the WPT phase, the WR-based WPT yields an average $v_\text{out}= 0.0512$\,V, less than $v_\text{out}= 0.0578$\,V achieved by the WS-based WPT.}, when $N_\text{FL} = 200 \text{ or } 300$.
4) As WR significantly reduces the overhead for searching a preferred waveform, the WR-based WPT can exploit a TS codebook, where the highest-level codewords have much higher resolution than the codewords in the codebook for the WS-based WPT. For instance, when $N_\text{FL} = 100$, the WS-based WPT is unable to exploit a codebook with $N_p = 2^9$, while the WR-based WPT can exploit a TS codebook with $L=9$.

\section{Conclusions}\label{SecConclu}
Relaxing the assumption of perfect Channel State Information at the Transmitter (CSIT), we have proposed novel waveform strategies for multi-antenna multi-sine Wireless Power Transfer (WPT) with limited feedback.
We respectively proposed the Waveform Selection (WS) and the Waveform Refinement (WR) to generate multi-sine signals for WPT from offline designed codebooks of waveform precoders.
These codebooks are optimized by proposing novel algorithms based on the framework of the generalized Lloyd's algorithm.
Simulation results show that the proposed waveform strategies, though relying on limited feedback, can outperform a variety of baselines, achieving a higher rectenna DC output voltage.
We have also drawn insights into the effect of various system parameters on the proposed waveform strategies.
The evaluations highlight that a large number of codewords is necessary for the WS-based WPT in the multi-antenna setup but leads to a long WS procedure, which is a high overhead. To alleviate the overhead, though the WR strategy has been proposed, a waveform strategy based on decoupled spatial beamforming and frequency domain precoding codebooks might be studied in future. The strategy relaying on decoupled codebooks may benefit the large-scale WPT with many antennas, reducing the overhead caused by determining preferred waveforms.

\appendix

\subsection{Proof of Theorem \ref{TheoAlgSAAconverge}}\label{AppTheoAlgSAAconverge}
\setcounter{equation}{0}
\renewcommand{\theequation}{A.\arabic{equation}}
As the optimized rank-1 {\small$\mathbf{X}^\star$} of problem (\ref{MaxVoutSAA_AP_Equiv}) is also optimal for problem (\ref{Epi_Prob_MaxVoutSAA_AP}), solving (\ref{MaxVoutSAA_AP_EquivQP}) essentially achieves the rank-constrained optimal {\small$\mathbf{X}^\star = \mathbf{x}^\star[\mathbf{x}^\star]^H$} of problem (\ref{Epi_Prob_MaxVoutSAA_AP}).
Then, we shall show that solving (\ref{Epi_Prob_MaxVoutSAA_AP}) iteratively yields a stationary point of problem (\ref{Epi_Prob_MaxVoutSAA}). As {\small$- g_q^{[t_0]}\big(\mathbf{t}_q^{[t_0]}\big)$} is convex, the first-order Taylor approximation {\small $\tilde{g}_q^{[t_0]}\big(\mathbf{t}_q^{[t_0]}, \big[\mathbf{t}_q^{[t_0]}\big]^{(l-1)} \big)$} satisfies {\small $g_q^{[t_0]}\big(\mathbf{t}_q^{[t_0]}\big) \leq \tilde{g}_q^{[t_0]} \big(\mathbf{t}_q^{[t_0]}, \big[\mathbf{t}_q^{[t_0]}\big]^{(l-1)}\big)$}. Then, {\small $\tilde{g}_q^{[t_0]} \big(\big[\mathbf{t}_q^{[t_0]}\big]^{(l)}, \big[\mathbf{t}_q^{[t_0]}\big]^{(l)} \big) = g_q^{[t_0]}\big( \big[\mathbf{t}_q^{[t_0]}\big]^{(l)} \big) \leq \tilde{g}_q^{[t_0]} \big( \big[\mathbf{t}_q^{[t_0]}\big]^{(l)}, \big[\mathbf{t}_q^{[t_0]}\big]^{(l-1)} \big)$}. In the above inequality, {\small$g_q^{[t_0]}\big( \big[\mathbf{t}_q^{[t_0]}\big]^{(l)} \big)$} basically connects the approximate problems (\ref{Epi_Prob_MaxVoutSAA_AP}) at iterations $(l-1)$ and $l$. The inequality indicates that the optimal solution of (\ref{Epi_Prob_MaxVoutSAA_AP}) at iteration {\small$(l-1)$} is always a feasible point of the approximate problem (\ref{Epi_Prob_MaxVoutSAA_AP}) at iteration {\small$l$}. As (\ref{Epi_Prob_MaxVoutSAA_AP}) is convex, its objective function {\small$\gamma_1^{(l)} \leq \gamma_1^{(l-1)}$}.
Given that the complex eigenvectors of {\small$\mathbf{A}_1$} can be uniquely attained, by using contradiction \cite[Proof of Proposition 2.7.1]{Bertsekas99}, we can show that the minimizer of (\ref{Epi_Prob_MaxVoutSAA_AP}) converges to a limit point. Hence, Algorithm \ref{AlgSAA} converges to a limit point. Therefore, as {\small$l \rightarrow \infty$}, {\small$\big[\mathbf{t}_q^{[t_0]}\big]^{(l)} \rightarrow \mathbf{\bar{t}}_q^{[t_0]}$}.
As {\small$\tilde{g}_q^{[t_0]} \big(\big[\mathbf{t}_q^{[t_0]}\big]^{(l)}, \big[\mathbf{t}_q^{[t_0]}\big]^{(l)} \big) = g_q^{[t_0]}\big( \big[\mathbf{t}_q^{[t_0]}\big]^{(l)} \big)$} and {\small$\nabla g_q^{[t_0]}\big( \big[\mathbf{t}_q^{[t_0]}\big]^{(l)} \big) = \nabla \tilde{g}_q^{[t_0]} \big( \big[\mathbf{t}_q^{[t_0]}\big]^{(l)}, \big[\mathbf{t}_q^{[t_0]}\big]^{(l)} \big)$}, it follows that {\small$g_q^{[t_0]}\big( \mathbf{\bar{t}}_q^{[t_0]} \big) = \tilde{g}_q^{[t_0]} \big(\mathbf{\bar{t}}_q^{[t_0]}, \mathbf{\bar{t}}_q^{[t_0]} \big)$} and {\small$\nabla g_q^{[t_0]}\big( \mathbf{\bar{t}}_q^{[t_0]} \big) =$ $ \nabla \tilde{g}_q^{[t_0]} \big( \mathbf{\bar{t}}_q^{[t_0]}, \mathbf{\bar{t}}_q^{[t_0]} \big)$}. Therefore, solving (\ref{Epi_Prob_MaxVoutSAA_AP}) iteratively yields a KKT point of problem (\ref{Epi_Prob_MaxVoutSAA_RkRelxd}). Then, we can conclude that the minimizer of problem (\ref{MaxVoutSAA_AP_Equiv}) converges to the KKT point of problem (\ref{Epi_Prob_MaxVoutSAA}). That is, designating the aforementioned limit point obtained by Algorithm \ref{AlgSAA} as $\mathbf{\bar{s}}$, $\mathbf{\bar{s}} \mathbf{\bar{s}}^H$ satisfies the KKT conditions of problem (\ref{Epi_Prob_MaxVoutSAA}).

\subsection{Proof of Proposition \ref{PropGLA_VQcdwdOpt}}\label{AppPropGLA_VQcdwdOpt}
\setcounter{equation}{0}
\renewcommand{\theequation}{B.\arabic{equation}}
In problem (\ref{Epi_Prob_MaxVoutSAA_AP}), as {\small$\tilde{g}_q^{[t_0]} \big(\big[\mathbf{t}_q^{[t_0]}\big]^{(l)}, \big[\mathbf{t}_q^{[t_0]}\big]^{(l)} \big) = g_q^{[t_0]}\big( \big[\mathbf{t}_q^{[t_0]}\big]^{(l)} \big)$}, {\small$\gamma_1^{(0)} = \sum_{t_0 = 1}^{T_0} \! \sum_{q = 1}^Q w_q \! \big(\! - \beta_2 \big[t_{q,0}^{[t_0]}\big]^{(0)} \!+\! \tilde{g}_q^{[t_0]}\! \big(\big[\mathbf{t}_q^{[t_0]}\big]^{(0)}\!, \big[\mathbf{t}_q^{[t_0]}\big]^{(0)}  \big)  \! \big) = \sum_{t_0 = 1}^{T_0} \! \sum_{q = 1}^Q w_q \! \big(\! - \beta_2 \big[t_{q,0}^{[t_0]}\big]^{(0)} \!+\! g_q^{[t_0]}\big( \big[\mathbf{t}_q^{[t_0]}\big]^{(0)} \big) \big)$}, where {\small$ \big[t_{q,k}^{[t_0]}\big]^{(0)} = \text{Tr}\Big\{ \mathbf{M}_{q,k}^{[t_0]} \mathbf{s}_\text{int} \mathbf{s}_\text{int}^H \Big\}$}.
As {\small $g_q^{[t_0]}\big(\mathbf{t}_q^{[t_0]}\big) \leq \tilde{g}_q^{[t_0]} \big(\mathbf{t}_q^{[t_0]}, \big[\mathbf{t}_q^{[t_0]}\big]^{(l-1)}\big)$}, {\small$\sum_{t_0 = 1}^{T_0} \! \sum_{q = 1}^Q w_q \! \big(\! - \beta_2 \big[t_{q,0}^{[t_0]}\big]^{(l)} \!+\! g_q^{[t_0]}\big( \big[\mathbf{t}_q^{[t_0]}\big]^{(l)} \big)  \big) \leq \sum_{t_0 = 1}^{T_0} \! \sum_{q = 1}^Q w_q \! \big(\! - \beta_2 \big[t_{q,0}^{[t_0]}\big]^{(l)} \!+\! \tilde{g}_q^{[t_0]}\! \big(\big[\mathbf{t}_q^{[t_0]}\big]^{(l)}\!, \big[\mathbf{t}_q^{[t_0]}\big]^{(l-1)}  \big)  \! \big) = \gamma_1^{(l)} $}.
According to Appendix \ref{AppTheoAlgSAAconverge}, {\small$\gamma_1^{(l)} \leq \gamma_1^{(l-1)}$}, such that {\small$\sum_{t_0 = 1}^{T_0} \! \sum_{q = 1}^Q w_q \! \big(\! - \beta_2 \big[t_{q,0}^{[t_0]}\big]^{(0)} \!+\! g_q^{[t_0]}\big( \big[\mathbf{t}_q^{[t_0]}\big]^{(0)} \big) \big) \geq \sum_{t_0 = 1}^{T_0} \! \sum_{q = 1}^Q w_q \! \big(\! - \beta_2 \big[t_{q,0}^{[t_0]}\big]^{(l)} \!+\! g_q^{[t_0]}\big( \big[\mathbf{t}_q^{[t_0]}\big]^{(l)} \big)  \big)$}. Hence, {\small $\bar{f}_d \big( \{\mathcal{H}_{0,n_p}^\prime\}_{n_p=1}^{N_p}, \{ \mathcal{S}_{\text{opt},n_p}^\prime \}_{n_p=1}^{N_p}, \mathcal{S} \big) \leq \bar{f}_d \big( \{\mathcal{H}_{0,n_p}^\prime\}_{n_p=1}^{N_p}, \{ \mathcal{S}_{\text{opt},n_p}^\prime \}_{n_p=1}^{N_p}, \mathcal{S}^\prime \big)$} provided that {\small$\mathbf{s}_\text{int} = [\mathcal{S^\prime}]_{n_p}$}.

\bibliographystyle{IEEEtran}
\bibliography{IEEEabrv,BibPro}

\begin{thebibliography}{10}
\providecommand{\url}[1]{#1}
\csname url@samestyle\endcsname
\providecommand{\newblock}{\relax}
\providecommand{\bibinfo}[2]{#2}
\providecommand{\BIBentrySTDinterwordspacing}{\spaceskip=0pt\relax}
\providecommand{\BIBentryALTinterwordstretchfactor}{4}
\providecommand{\BIBentryALTinterwordspacing}{\spaceskip=\fontdimen2\font plus
\BIBentryALTinterwordstretchfactor\fontdimen3\font minus
  \fontdimen4\font\relax}
\providecommand{\BIBforeignlanguage}[2]{{%
\expandafter\ifx\csname l@#1\endcsname\relax
\typeout{** WARNING: IEEEtran.bst: No hyphenation pattern has been}%
\typeout{** loaded for the language `#1'. Using the pattern for}%
\typeout{** the default language instead.}%
\else
\language=\csname l@#1\endcsname
\fi
#2}}
\providecommand{\BIBdecl}{\relax}
\BIBdecl

\bibitem{ZCZ16}
Y.~Zeng, B.~Clerckx, and R.~Zhang, ``Communications and signals design for
  wireless power transmission,'' \emph{{IEEE} Trans. Commun.}, vol.~65, no.~5,
  pp. 2264--2290, May 2017.

\bibitem{BBFCGC15}
A.~Boaventura~{\sl et al}., ``Boosting the efficiency: {U}nconventional
  waveform design for efficient wireless power transfer,'' \emph{{IEEE} Microw.
  Mag.}, vol.~16, no.~3, pp. 87--96, Apr. 2015.

\bibitem{CM16}
A.~Costanzo and D.~Masotti, ``Smart solutions in smart spaces: {Getting} the
  most from far-field wireless power transfer,'' \emph{{IEEE} Microw. Mag.},
  vol.~17, no.~5, pp. 30--45, May 2016.

\bibitem{BCCG13}
A.~Boaventura, A.~Collado, N.~B. Carvalho, and A.~Georgiadis, ``Optimum
  behavior: {Wireless} power transmission system design through behavioral
  models and efficient synthesis techniques,'' \emph{{IEEE} Microw. Mag.},
  vol.~14, no.~2, pp. 26--35, Mar. 2013.

\bibitem{BC11}
A.~Boaventura and N.~Carvalho, ``Maximizing dc power in energy harvesting
  circuits using multisine excitation,'' in \emph{2011 IEEE MTT-S International
  Microwave Symposium Digest}, Jun. 2011.

\bibitem{CB16}
B.~Clerckx and E.~Bayguzina, ``Waveform design for wireless power transfer,''
  \emph{{IEEE} Trans. Signal Process.}, vol.~64, no.~23, pp. 6313--6328, Dec.
  2016.

\bibitem{HC16SPAWC}
Y.~Huang and B.~Clerckx, ``Waveform optimization for large-scale multi-antenna
  multi-sine wireless power transfer,'' in \emph{IEEE International Workshop on
  Signal Processing Advances in Wireless Communications 2015}, Jul. 2016.

\bibitem{HC16TSParxiv}
------, ``Large-scale multiantenna multisine wireless power transfer,''
  \emph{{IEEE} Trans. Signal Process.}, vol.~65, no.~21, pp. 5812--5827, Nov.
  2017.

\bibitem{CB17arXiv}
B.~Clerckx and E.~Bayguzina, ``A low-complexity adaptive multisine waveform
  design for wireless power transfer,'' \emph{{IEEE} Antennas Wireless Propag.
  Lett.}, vol.~16, pp. 2207--2210, May 2017.

\bibitem{MZZ17arxiv}
M.~R.~V. Moghadam, Y.~Zeng, and R.~Zhang, ``Waveform optimization for
  radio-frequency wireless power transfer,'' submitted for publication,
  available online at arXiv:1703.04006.

\bibitem{ZZH13}
X.~Zhou, R.~Zhang, and C.~K. Ho, ``Wireless information and power transfer:
  {Architecture} design and rate-energy tradeoff,'' \emph{{IEEE} Trans.
  Commun.}, vol.~61, no.~11, pp. 4754--4767, Oct. 2013.

\bibitem{LBG80}
Y.~Linde, A.~Buzo, and R.~Gray, ``An algorithm for vector quantizer design,''
  \emph{{IEEE} Trans. Commun.}, vol.~28, no.~1, pp. 84--95, Jan. 1980.

\bibitem{XG06}
P.~Xia and G.~B. Giannakis, ``Design and analysis of transmit-beamforming based
  on limited-rate feedback,'' \emph{{IEEE} Trans. Signal Process.}, vol.~54,
  no.~5, pp. 1853--1863, May 2006.

\bibitem{KJL12}
K.~Ko, S.~Jung, and J.~Lee, ``Hybrid {MU-MISO} scheduling with limited feedback
  using hierarchical codebooks,'' \emph{{IEEE} Trans. Commun.}, vol.~60, no.~4,
  pp. 1101--1113, Apr. 2012.

\bibitem{JWSWSC12}
C.~Jiang, M.~M. Wang, F.~Shu, J.~Wang, W.~Sheng, and Q.~Chen, ``Multi-user mimo
  with limited feedback using alternating codebooks,'' \emph{{IEEE} Trans.
  Commun.}, vol.~60, no.~2, pp. 333--338, Feb. 2012.

\bibitem{CKK08}
B.~Clerckx, G.~Kim, and S.~Kim, ``Correlated fading in broadcast {MIMO}
  channels: {Curse} or blessing?'' in \emph{2008 IEEE Global Telecommunications
  Conference (GLOBCOM)}, Nov. 2008, pp. 1--5.

\bibitem{CO13}
B.~Clerckx and C.~Oestges, \emph{{MIMO} Wireless Networks: Channels, Techniques
  and Standards for Multi-Antenna, Multi-User and Multi-Cell Systems},
  2nd~ed.\hskip 1em plus 0.5em minus 0.4em\relax Academic Press, 2013.

\bibitem{SM11}
W.~Santipach and K.~Mamat, ``Tree-structured random vector quantization for
  limited-feedback wireless channels,'' \emph{{IEEE} Trans. Wireless Commun.},
  vol.~10, no.~9, pp. 3012--3019, Sep. 2011.

\bibitem{YHG14}
G.~Yang, C.~K. Ho, and Y.~L. Guan, ``Dynamic resource allocation for
  multiple-antenna wireless power transfer,'' \emph{{IEEE} Trans. Signal
  Process.}, vol.~62, no.~14, pp. 3565--3577, Jul. 2014.

\bibitem{ZZ15}
Y.~Zeng and R.~Zhang, ``Optimized training design for wireless energy
  transfer,'' \emph{{IEEE} Trans. Commun.}, vol.~63, no.~2, pp. 536--550, Feb.
  2015.

\bibitem{XZ14}
J.~Xu and R.~Zhang, ``Energy beamforming with one-bit feedback,'' \emph{{IEEE}
  Trans. Signal Process.}, vol.~62, no.~20, pp. 5370--5381, Oct. 2014.

\bibitem{CKC17}
K.~W. Choi, D.~I. Kim, and M.~Y. Chung, ``Received power-based channel
  estimation for energy beamforming in multiple-antenna {RF} energy transfer
  system,'' \emph{{IEEE} Trans. Signal Process.}, vol.~65, no.~6, pp.
  1461--1476, Mar. 2017.

\bibitem{XZ16}
J.~Xu and R.~Zhang, ``A general design framework for {MIMO} wireless energy
  transfer with limited feedback,'' \emph{{IEEE} Trans. Signal Process.},
  vol.~64, no.~10, pp. 2475--2488, May 2016.

\bibitem{ZZjun15}
Y.~Zeng and R.~Zhang, ``Optimized training for net energy maximization in
  multi-aantenna wireless energy transfer over frequency-selective channel,''
  \emph{{IEEE} Trans. Commun.}, vol.~63, no.~6, pp. 2360--2373, Jun. 2015.

\bibitem{OCV11}
U.~Olgun, C.~C. Chen, and J.~L. Volakis, ``Investigation of rectenna array
  configurations for enhanced {RF} power harvesting,'' \emph{{IEEE} Antennas
  Wireless Propag. Lett.}, vol.~10, pp. 262--265, Apr. 2011.

\bibitem{Wetenkamp83}
S.~Wetenkamp, ``Comparison of single diode vs. dual diode detectors for
  microwave power detection,'' in \emph{1983 IEEE MTT-S International Microwave
  Symposium Digest}, May 1983, pp. 361--363.

\bibitem{SDR09}
A.~Shapiro, D.~Dentcheva, and A.~Ruszcynski, \emph{Lectures on Stochastic
  Programming: {M}odeling and Theory}.\hskip 1em plus 0.5em minus 0.4em\relax
  Society for Industrial and Applied Mathematics Philadelphia, 2009.

\bibitem{HP10}
Y.~Huang and D.~P. Palomar, ``Rank-constrained separable semidefinite
  programming with applications to optimal beamforming,'' \emph{{IEEE} Trans.
  Signal Process.}, vol.~58, no.~2, pp. 664--678, Feb. 2010.

\bibitem{GB14}
M.~Grant and S.~Boyd, ``{CVX}: {Matlab} software for disciplined convex
  programming, version 2.1,'' \url{http://cvxr.com/cvx}, Mar. 2014.

\bibitem{HP14}
Y.~Huang and D.~P. Palomar, ``Randomized algorithms for optimal solutions of
  double-sided {QCQP} with applications in signal processing,'' \emph{{IEEE}
  Trans. Signal Process.}, vol.~62, no.~5, pp. 1093--1108, Mar. 2014.

\bibitem{BN01}
A.~Ben-Tal and A.~Nemirovski, \emph{Lectures on Modern Convex Optimization:
  Analysis, Algorithms, and Engineering Applications}, ser. MPSSIAM {Series} on
  Optimization.

\bibitem{Parlett00}
B.~N. Parlett, ``The {QR} algorithm,'' \emph{{IEEE} Comput. Sci. Eng.}, vol.~2,
  no.~1, pp. 38--42, Jan. 2000.

\bibitem{GG92}
A.~Gersho and R.~M. Gray, \emph{Vector Quantization and Signal
  Compression}.\hskip 1em plus 0.5em minus 0.4em\relax Boston, MA: Kluwer,
  1992.

\bibitem{BR14}
C.~Boyer and S.~Roy, ``Backscatter communication and {RFID}: {Coding}, energy,
  and {MIMO} analysis,'' \emph{{IEEE} Trans. Commun.}, vol.~62, no.~3, pp.
  770--785, Mar. 2014.

\bibitem{ESetal04}
V.~Erceg~{\ et al}, ``{TGn} channel models,'' in \emph{Version 4. IEEE
  802.11-03/940r4}, May 2004.

\bibitem{KCM17}
J.~Kim, B.~Clerckx, and P.~D. Mitcheson, ``Prototyping and experimentation of a
  closed-loop wireless power transmission with channel acquisition and waveform
  optimization,'' in \emph{2017 IEEE WPTC}, May 2017, pp. 1--4.

\bibitem{Bertsekas99}
D.~P. Bertsekas, \emph{Nonlinear Programming}, 2nd~ed.\hskip 1em plus 0.5em
  minus 0.4em\relax Belmont, MA: Athena Scientific, 1999.

\end{thebibliography}

\end{document}